# HL-nets: Physics-informed neural networks for hydrodynamic lubrication with cavitation

Yiqian Cheng[†], Qiang He[†], Weifeng Huang[*], Ying Liu, Yanwen Li, Decai Li

*State Key Laboratory of Tribology in Advanced Equipment, Department of Mechanical Engineering, Tsinghua University, Beijing 100084, China*

**Abstract**

Recently, physics-informed neural networks (PINNs) have emerged as a promising method for solving partial differential equations (PDEs). In this study, we establish a deep learning computational framework, HL-nets, for computing the flow field of hydrodynamic lubrication involving cavitation effects. Two classical cavitation conditions, i.e., the Swift-Stieber (SS) condition and the Jakobsson-Floberg-Olsson (JFO) condition, are implemented in the PINNs to solve the Reynolds equation. For the non-negativity constraint of the SS cavitation condition, a penalizing scheme with a residual of the non-negativity and an imposing scheme with a continuous differentiable non-negative function are proposed. For the complementarity constraint of the JFO cavitation condition, the pressure and cavitation fraction are taken as the neural network outputs, and the residual of the Fischer-Burmeister (FB) equation constrains their complementary relationships. Multi-task learning (MTL) methods are applied to balance the newly introduced loss terms described above. To estimate the accuracy of HL-nets, we present a numerical solution of the Reynolds equation for oil-lubricated bearings involving cavitation. The results indicate that the proposed HL-nets can highly accurately simulate hydrodynamic lubrication involving cavitation phenomena. The imposing scheme can effectively improve the accuracy of the training results of PINNs, and it is expected to have great potential to be applied to different fields where the non-negativity constraint is needed.

## 1. Introduction

Friction, which occurs when solid surfaces are in contact with one another and moving toward each other, causes energy loss and wear of mechanical systems; so, reducing friction is essential to increasing the cost efficacy of energy systems. To reduce friction and wear, hydrodynamic lubrication is widely used to form a thin lubricant film between solid surfaces with normal load-bearing capacity and low shear strength [1]. As a result of Reynolds' work [2], the Reynolds equation was developed, which is a elliptic PDE governing the pressure distribution of thin viscous fluid films; as one of the fundamental equations of the classical lubrication theory, the Reynolds equation has been extensively applied to a wide range of lubrication problems with indisputable success[3,4].

[†] These authors contributed equally to this work.
[*] Corresponding author.
E-mail address: huangwf@tsinghua.edu.cn (W.Huang)

Cavitation in liquid lubricating films is common. When surfaces covered in hydrodynamic lubrication contain diverging regions, the pressure will drop, and cavitation will occur when the pressure drops to a certain level [5]. Cavitation directly impacts the lubrication film's pressure distribution, affecting the load-carrying capacity and friction. Thus, cavitation phenomena play a crucial role in lubrication modeling. While liquids can be maintained at a certain pressure and the cavitation region contains a mixture of gas and vapor, the details of cavitation are still an area of interest to researchers studying bubble dynamics and cavitation theory [6]. In the domain of tribology, it is usually assumed that a liquid will completely vaporize when the pressure of the liquid in a region is below zero or a constant cavitation pressure. This restriction on the pressure field by introducing cavitation into the Reynolds equation is called the cavitation condition.

Over the past few decades, several mathematical models and numerical methods have been developed to address the cavitation problem, among which the most important models are the Swift-Stieber (SS) condition and the Jakobsson-Floberg-Olsson (JFO) condition. The SS cavitation condition, formulated independently by Swift [7] and Stieber and Schwimmlager [8], has been widely used because of its simplicity, ease of implementation, and superior accuracy compared to the full- and half-Sommerfeld conditions. The SS cavitation model is a typical obstacle problem [9]. This condition assumes that the pressure in the cavitation region is the cavitation pressure, and the pressure gradient is zero at the boundary of the cavitation region. However, this boundary condition does not include the film reformation boundary where the cavitation ends and the full film begins; hence, it does not enforce mass conservation. The JFO model for cavitation was proposed [10,11] to account for film reformation and to ensure mass conservation. In the tribology field, this model is known as the JFO boundary conditions. By incorporating the binary switch function, Elrod [12] provided the first algorithm to incorporate the JFO boundary conditions in a single equation for both full-film and cavitated regions, which can predict the cavitation and full-film regions. Several modifications of the Elrod algorithm have been proposed to improve convergence for the highly nonlinear nature of the algorithm with a binary switch function [13–17]. Upon considering mass-conserving cavitation, the governing equation remains elliptic within full-film regions but becomes hyperbolic within cavitated regions, forming a mixture of nonlinear PDEs. For numerical stability and accuracy, the convective terms are discretized by the finite difference method and finite volume method using the upwind format [18–21], and the weak format in the finite element method (FEM) requires a stabilization method, such as the Streamline Upwind/Petrov-Galerkin (SUPG) method [22]. Giacopini et al. [23] converted the cavitation problem into a linear-complementarity problem by constructing a complementarity condition between pressure and cavitation fraction. Woloszynski et al. [24] developed an efficient algorithm, Fischer-BurmeisterNewton-Schur, and reformulated the constrained optimization problem into an unconstrained one. The two parameters, pressure and cavitation fraction, were determined simultaneously by a system of nonlinear equations formed by discretizing the Reynolds equation and the Fischer-Burmeister (FB) complementary equation, respectively. The method has sufficient accuracy and very high efficiency with low-cost gradient-based methods. This complementary function approach has also been combined with many other numerical algorithms [25–27].

The above-mentioned studies on the Reynolds equation and cavitation are based on the FEM, finite volume method, finite-difference method, or other traditional numerical methods. However, a new deep learning method called physics-informed neural networks (PINNs) has emerged recently as a promising method for solving forward and inverse PDEs. Raissi et al. [28] introduced a general framework of PINNs and verified its ability to solve PDEs and their inverse problems, such as extracting parameters from a few observations. Compared to other deep learning approaches in physical modeling, the PINNs approach incorporates physical laws and PDEs in addition to using data in a flexible manner. The PINNs are not subject to Courant-Friedrichs-Lewy (CFL) constraints for transient equations, and the time discretization terms can be treated as general boundary conditions [29]. For equations containing convective terms, the flow field direction is generally not treated during the solution of PINNs without additional treatment [30]. PINNs have been extensively applied in recent years to forward and inverse problems in fluid mechanics [31–35], solid mechanics [36,37], heat transfer [38,39], and flow in porous media [40] [41]. Numerous function libraries have been developed for PINNs to make it easier for researchers to use [42–44].

Some preliminary studies on the solution of PINNs for hydrodynamic lubrication have been published recently. For example, Almqvist [45] and Zhao et al. [46] used PINNs to solve the one- and two-dimensional steady-state Reynolds equation for a linear slider. Li et al. [47] devised a PINN scheme to solve the Reynolds equation to predict the gas bearing's flow fields and aerodynamic characteristics. Despite being an effective numerical tool, PINNs have not been well applied in the field of tribology. The only exploratory studies on PINN methods have been based on the most basic PINNs, and some newer approaches such as adaptive methods and special treatment of boundary conditions have not been discussed. In terms of tribological problems, the solution of the Reynolds equation is still limited to specific cases of cavitation-free applications such as linear slider and gas bearing, while more practical applications of hydrodynamic lubrication problems involving cavitation (such as oil- or water-lubricated bearings) still cannot be solved by PINNs. The SS cavitation condition makes hydrodynamic lubrication become a typical obstacle problem; in terms of the JFO cavitation condition, the cavitation fraction has a jump condition, which makes it challenging for the PINNs to solve. None of these issues have been addressed to enable the use of PINNs for simulating hydrodynamic lubrication.

In this study, HL-nets, a PINNs-based solver, is developed to solve the Reynolds equation, and the SS cavitation and JFO cavitation conditions are introduced into the PINNs to ensure HL-nets is applicable to simulating hydrodynamic lubrication with cavitation. In addition to the penalizing scheme of traditional PINNs, an imposing scheme is developed to imply different conditions, including the Dirichlet boundary (DB) condition, SS cavitation condition, and JFO cavitation condition. Multi-task learning (MTL) methods are applied to balance the newly introduced loss terms described above. The oil-lubricated bearing is studied as a typical hydrodynamic lubrication problem to test the performance of the proposed HL-nets; the traditional penalizing schemes and imposing condition schemes for DB conditions are compared, and cavitation problems with SS or JFO cavitation conditions are studied.

The rest of the paper is organized as follows: in Section 2, the mathematical modeling of HL-nets is presented, including an introduction to the Reynolds equation and several cavitation conditions as well as descriptions of PINNs designed for solving the Reynolds equation with the cavitation conditions. Numerical tests of oil-lubricated bearings are performed to validate HL-nets in Section 3. Some concluding remarks and a brief discussion are presented in Section 4.

## 2. Methodology

### 2.1 Reynolds equation and cavitation conditions

The Reynolds equation describes the flow of a thin lubricant film between two surfaces. This equation was derived from the N-S equation based on several assumptions, including ignoring inertial forces and pressure changes along the thickness direction of the lubrication film. For a Newtonian fluid with constant viscosity, the steady-state Reynolds equation can be expressed as–

$$\nabla \cdot (\rho h^3 \nabla p) = 6\mu \boldsymbol{U} \cdot \nabla(\rho h), \tag{1}$$

where $p$ is the pressure, $h$ is the film thickness, $\mu$ is the viscosity of the fluid, $\rho$ is the density of the fluid, and $\boldsymbol{U}$ is the relative sliding velocity. Without loss of generality, we can assume that the sliding velocity is always directed along the $x$-axis direction. Then, Eq. (1) can be rewritten as

$$\nabla \cdot (\rho h^3 \nabla p) = 6\mu U \frac{\partial(\rho h)}{\partial x}. \tag{2}$$

- DB condition

The pressure at the boundary is always specified as the ambient pressure $p_{\partial\Omega}$ for the hydrodynamic lubrication problem, which is the DB condition for the Reynolds equation:

$$p = p_{\partial\Omega}. \tag{3}$$

The Reynolds equation (Eq. (2)) can be solved with the DB condition (Eq. (3)), but there is a possibility that the pressure value might be smaller than the cavitation pressure $p_{cav}$ in applications where there is an evanescent gap (e.g., sliding bearings and thrust bearings with surface textures). To bridge the gap between simulation and physical reality, two different cavitation conditions have been developed based on the Reynolds equation, namely the SS cavitation condition and the JFO cavitation condition.

- SS cavitation condition

The cavitation pressure $p_{cav}$ is set to zero for simplicity in this study, which can be derived directly by translational transformation. Then, the SS cavitation condition imposes the following pressure constraints in the cavitation region [7,8]:

$$\nabla p = 0, p = p_{cav} = 0, \tag{4}$$

which indicates that the pressure is continuous and differentiable throughout the flow field, and it is subject to a physical lower limit. This is a typical obstacle problem [9], and the governing equations in the form of a variational inequality are

$$\begin{cases} -\nabla \cdot (\rho h^3 \nabla p) \geq -6\mu U \frac{\partial(\rho h)}{\partial x}, \\ p \geq 0, \\ \left[\nabla \cdot (\rho h^3 \nabla p) - 6\mu U \frac{\partial(\rho h)}{\partial x}\right] p = 0. \end{cases} \tag{5}$$

- JFO cavitation condition

In the JFO cavitation model, the fluid region consists of a full-film region and a cavitation region with varying fluid densities. The cavitation fraction $\theta$ assumes two extreme values, zero and one, in the bulk of the full-film region and cavitation region. The density of the lubricant film $\rho$ is a function of the cavitation fraction $\theta$ and the constant reference density of the lubricant $\rho_0$:

$$\rho = \rho_0(1 - \theta), \tag{6}$$

where $\rho_0$ is the constant density in the full-film region. Substituting Eq. (6) into Eq. (2) yields–

$$\nabla \cdot (h^3 \nabla p) = 6\mu U \frac{\partial[(1-\theta)h]}{\partial x}. \tag{7}$$

In the full-film region, Eq. (7) can be reduced to an incompressible fluid Reynolds equation. The JFO cavitation model can be converted into a complementary problem:

$$\nabla \cdot (h^3 \nabla p) = 6\mu U \frac{\partial[(1-\theta)h]}{\partial x}, \quad p\theta = 0, \quad p >= 0, \quad \theta >= 0. \tag{8}$$

The complementary relationship between pressure and cavitation fraction can be constrained by introducing the FB equation:

$$p + \theta - \sqrt{p^2 + \theta^2} = 0, \tag{9}$$

which satisfies the non-negativity of the pressure.

To improve the numerical stability of the calculation in practice, the above equations need to be dimensionalized to obtain

$$\frac{\partial}{\partial X}\left(H^3 \frac{\partial P}{\partial X}\right) + \frac{L^2}{B^2} \frac{\partial}{\partial Y}\left(H^3 \frac{\partial P}{\partial Y}\right) = \frac{\partial[(1-\theta)H]}{\partial X}, \tag{10}$$

$$X = \frac{x}{L}; Y = \frac{y}{B}; H = \frac{h}{h_0}; P = \frac{h_0^2}{6\eta UL}p, L = 2\pi R,$$

where $L$ and $B$ represent the length and width of the lubrication region, respectively, $p_{cav}$ represents the cavitation pressure, $h_0$ represents the minimum film thickness, and $R$ represents the radius of bearings. It should be noted that Eq. (10) can be viewed as a generalized form of the Reynolds equation, where the cavitation fraction $\theta$ is set to zero if the JFO cavitation condition is not used.

## 2.2 PINNs for solving Reynolds equation in HL-nets

### 2.2.1 PINNs for solving Reynolds equation

A scheme of the PINN framework is depicted in Fig. 2. In this scheme, $u(x)$ is approximated by a feed-forward fully connected network, for which the $x$ coordinate is the input and $P_\Theta$ is the output, where $\Theta$ denotes the trainable parameter set of the neural network. In this study, $u(x)$ can be the pressure $P(x)$ or the cavitation fraction $\theta(x)$.

The fully connected neural network is used to approximate the solution $P(x)$. As shown in Fig. 1, the neural network is composed of an input layer, $M-1$ hidden layers, and an output layer, as follows:

**Input layer:** $\quad \widetilde{u}^{[0]} = x,$

**Hidden layers:** $\quad \widetilde{u}^{[m]} = \sigma(W^{[m]}\widetilde{u}^{[m-1]} + b^{[m]})$, for $m = 2, 3, \dots, M-1,$ $\tag{11}$

**Output layer:** $\quad P_\Theta = \widetilde{u}^{[M]} = W^{[M]}\widetilde{u}^{[M-1]} + b^{[M]},$

where $\sigma(.)$ is the activation function representing a simple nonlinear transformation, such as $relu(.)$, $softmax(.)$, or $simgoid(.)$, and $W^{[m]}$ and $b^{[m]}$ are trainable weights and biases, respectively, at the $m$-th layer. All the trainable weights and biases form the trainable parameter set of the neural network, $\Theta = \{W^{[m]}, b^{[m]}\}_{1 \leq m \leq M}$.

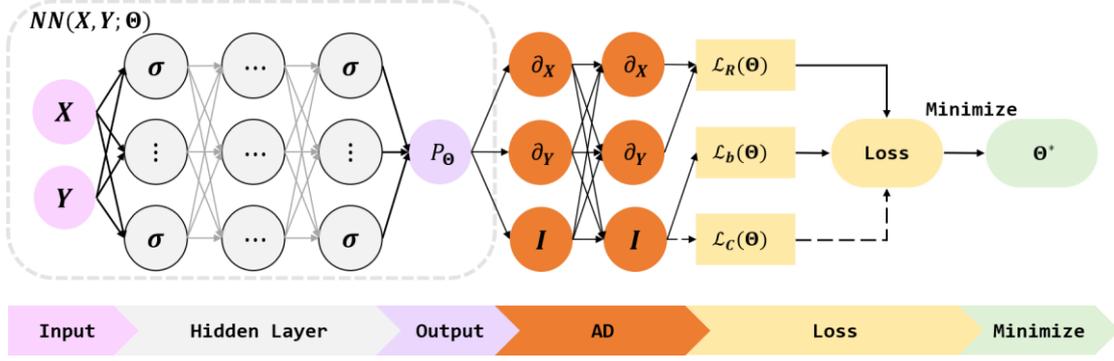

Fig. 1 General structure of PINNs for solving Reynolds equation.

The traditional scheme of PINNs for solving the Reynolds equation involves defining the loss of the Reynolds equation $\mathcal{L}_R(\Theta)$ and the loss of the constraint conditions $\mathcal{L}_i(\Theta)$. $P(x)$ is the solution of the Reynolds equation, which is approximated by the PINN output $P_\Theta(x)$. The parameters of the network in the traditional scheme can be trained by minimizing a composite loss function taking the form

$$\mathcal{L}(\Theta) := \lambda_R \mathcal{L}_R(\Theta) + \sum_{i=1}^{N_c} \lambda_i \mathcal{L}_i(\Theta). \qquad (12)$$

The mean square error is adopted to measure the loss of the neural network; the Reynolds equation loss $\mathcal{L}_R(\Theta)$ can be defined as

$$\mathcal{L}_R(\Theta) = \frac{1}{N_R} \sum_{i=1}^{N_R} \left[ \frac{\partial}{\partial X}\left( H^3(x^i) \frac{\partial P_\Theta(x^i)}{\partial X} \right) + \frac{L^2}{B^2} \frac{\partial}{\partial Y}\left( H^3(x^i) \frac{\partial P_\Theta(x^i)}{\partial Y} \right) - \frac{\partial \left[ \left(1-\theta(x^i)\right) H(x^i) \right]}{\partial X} \right]^2, \qquad (13)$$

where $N_R$ is the number of data points for the bulk domain and $\lambda_R$ and $\lambda_i$ are weight parameters used to balance different loss terms. It should be noted that the cavitation fraction $\theta(x^i)$ is set to zero, unless the JFO cavitation condition is used. The constraint condition loss $\mathcal{L}_i(\Theta)$ is different for the DB, SS, and JFO cavitation conditions; so, the detailed expression is not given here.

Calculating the residuals of Eq. (13) requires derivations of the outputs with respect to the inputs, which can be achieved conveniently by automatic differentiation. Automatic differentiation capabilities are widespread in deep-learning frameworks such as TensorFlow [48] and PyTorch [49]. Using the automatic differentiation method can eliminate the need to perform tedious derivations or numerical discretizations to calculate the derivatives in space and time. The parameters of the fully connected network are trained using gradient-descent methods based on the backpropagation of the loss function as follows:

$$\Theta^{(k+1)} = \Theta^{(k)} - \eta \nabla_\Theta \mathcal{L}\left(\Theta^{(k)}\right), \qquad (14)$$

where $\eta$ is the learning rate, and $k$ is the iteration step.

An important part of solving a PDE is to apply boundary conditions or other constraint conditions. Using a soft constraint for the penalizing scheme is the traditional method in PINNs. As expressed in Eq. (12), the total loss $\mathcal{L}(\Theta)$ is composed of different kinds of loss, and the constraint condition is achieved by reducing the residual between the numerical value and theoretical value. Although this method is effective for training PINNs of boundary conditions and other constraint conditions, the softly penalizing constraining scheme does not ensure that the constraint conditions are strictly satisfied, resulting in computational errors. Recently, the method of embedding the constraint conditions into the network structure was proposed [50,51]. Sukumar et al. [50] introduced an approach based on distance fields to exactly impose constraint conditions in PINNs, including Dirichlet, Neumann, and Robin boundary conditions. In their method, a distance function is multiplied by the output of the neural network and then added to the value required by the constraint condition, the result of which is used as the output of PINNs; the formula for imposing constraint conditions is

$$\boldsymbol{P_\Theta} = \varphi(\tilde{\boldsymbol{u}}^{[M]}), \tag{15}$$

where $M$ denotes the last layer of the neural network, and $\varphi$ is a function used to apply the constraint condition, which can be the DB or SS cavitation condition. When the imposing method is adopted for a constraint condition or a boundary condition, the corresponding loss term for this condition in Eq. (5) can be canceled.

**2.2.2 Multi-task learning (MTL)**

The optimization direction and optimization efficiency of the network are directly affected by loss weight selection. The total loss expression in Eq. (12) can be rewritten into a more general form:

$$\mathcal{L}(\boldsymbol{\Theta}) \coloneqq \sum_i^{N_t} \lambda_i \mathcal{L}_i(\boldsymbol{\Theta}), \tag{16}$$

where $N_t$ is the number of loss terms. With a large weight coefficient, the corresponding term contributes more to the total loss, and then, an optimization process is carried out to reduce this loss term. Therefore, it is important to choose reasonable weight coefficients.

For classical PINNs with only PDE loss and boundary loss, a too-large or too-small weighting factor of the boundary residuals will make the computational results less accurate [52,53]. It is possible to search for suitable multi-losses weight parameters via manual scaling to make the results of PINNs highly accurate, but this approach is very costly and tedious. Balancing different loss terms during training is typical of MTL, a learning technique that allows multiple tasks to be tackled concurrently based on shared representations [54]. Three different MTL methods are adopted in this study, namely the dynamic weight (DW) method, the uncertainty weight (UW) method, and the projecting conflicting gradient (PCGrad) method.

- DW method

The DW method [53] is widely used for balancing loss of governing equations and boundary conditions in the field of PINNs [55]. The main idea of DW is balancing the gradients of each different loss term to the same order of magnitude during PINN training. The weights are updated as

$$\hat{\lambda}_i^{(k+1)} = \frac{\overline{|\nabla_\Theta \mathcal{L}_R(\boldsymbol{\Theta})|}}{|\nabla_\Theta \mathcal{L}_i(\boldsymbol{\Theta})|}, \tag{17}$$

$$\lambda_i^{(k+1)} = (1-\alpha)\lambda_i^{(k)} + \alpha \hat{\lambda}_i^{(k+1)},$$

where $\alpha$ is a hyperparameter determining how fast the contributions of previous dynamic weights $\lambda_i^{(k)}$ decay, and it is set to 0.1 in this study, and $k$ is the training step.

- UW method

The UW method uses homoscedastic uncertainty to balance the multi-task losses based on maximizing the Gaussian likelihood [56]. When a task's homoscedastic uncertainty is high, the effect of this task on the network weight update is small. The loss function is defined as

$$\mathcal{L}(\mathbf{\Theta}; \mathbf{s}) := \sum_{i=1}^{N_T} \left(\frac{1}{2}\exp(-s_i)\mathcal{L}_i(\mathbf{\Theta}) + s_i\right), \tag{18}$$

where the log variance $s_i$ balances the task-specific losses during training [57], and $N_T$ is the number of loss terms. The goal is to find the best model weights $\mathbf{\Theta}$ and log variance $\mathbf{s}$ by minimizing the loss $\mathcal{L}(\mathbf{\Theta}; \mathbf{s})$. It should be noted that the training results of the neural network may be affected by the initialization of the log variance $s_i$.

- PCGrad method

The PCGrad method is a form of gradient surgery that projects a task's gradient onto the normal plane of the gradient of any other task with a conflicting gradient [58]. The PCGrad method is applied to the training of the neural network rather than applying constraint conditions directly, which is applied in the solution of PINNs and proved to be effective [59]. The parameters of the neural network are trained as follows:

$$\mathbf{\Theta}^{(k+1)} = \mathbf{\Theta}^{(k)} - \eta \cdot \text{PCGrad}(\{\nabla_{\mathbf{\Theta}}[\mathcal{L}_i(\mathbf{\Theta}^{(k)})]\}). \tag{19}$$

In a general form of the loss function with multiple loss terms, all training processes of the MTL method mentioned above are denoted as

$$\mathbf{\Theta}^{(k+1)} = \mathbf{\Theta}^{(k)} - \eta \cdot \text{MTL}(\{\nabla_{\mathbf{\Theta}}[\mathcal{L}_i(\mathbf{\Theta}^{(k)})]\}), \tag{20}$$

where $\text{MTL}(\cdot)$ represents an MTL method, specifically, $\text{DW}(\cdot)$, $\text{UW}(\cdot)$, or $\text{PCGrad}(\cdot)$ in this study.

## 2.3 PINNs for cavitation conditions in HL-nets

There are three different constraint conditions in HL-nets, namely the DB condition, SS cavitation condition, and JFO cavitation condition. As described above, there are two different schemes to apply constraint conditions—the penalizing scheme and the imposing scheme. In this section, the implementation of the two schemes for three different constraint conditions is described.

**2.3.1 PINNs for solving Reynolds equation with Dirichlet boundary condition**

Satisfying the DB condition is the basis for solving the Reynolds equation in hydrodynamic lubrication. It should be noted that when applying SS or JFO cavitation conditions, the DB condition should also be satisfied at the boundary of the computational domain. For simplicity, we denote the penalizing scheme for the DB condition as the "Pe-DB" scheme and the imposing scheme for the DB condition as the "Imp-DB" scheme.

- Pe-DB scheme

The DB condition loss $\mathcal{L}_b(\mathbf{\Theta})$ is defined as

$$\mathcal{L}_b(\mathbf{\Theta}) = \frac{1}{N_b}\sum_{i=1}^{N_b}[\mathbf{P}_{\mathbf{\Theta}}(\mathbf{x}^i) - P_{\partial\Omega}(\mathbf{x}^i)]^2. \tag{21}$$

The parameters of the fully connected network are trained based on the backpropagation of the loss function consisting of the Reynolds equation loss $\mathcal{L}_R(\mathbf{\Theta})$ and DB condition loss $\mathcal{L}_b(\mathbf{\Theta})$. The multiple loss terms in the Pe-DB scheme are denoted as

$$\mathcal{L}(\mathbf{\Theta}) = [\mathcal{L}_R(\mathbf{\Theta}), \mathcal{L}_b(\mathbf{\Theta})]. \tag{22}$$

- Imp-DB scheme

An approximate distance function $\phi$ to the boundary of a domain is multiplied by the output of the neural network and then added to the boundary value function $\mathbf{P}_{\partial\Omega}$ required by the boundary condition, the result of which is used as the output of PINNs [50]. Elimination of one loss will

significantly improve the accuracy and stability of the calculation. The specific formula is–
$$\boldsymbol{P}_{\boldsymbol{\Theta}} = \varphi(\widetilde{\boldsymbol{P}}^{[M]}) = \boldsymbol{P}_{\partial\Omega} + \phi\widetilde{\boldsymbol{P}}^{[M]}, \qquad (23)$$
where $\phi$ is the distance function with a value of zero at boundaries. In this scheme, only Reynolds equation residuals remain in the loss function; so, the loss term is denoted as
$$\mathcal{L}(\boldsymbol{\Theta}) = [\mathcal{L}_R(\boldsymbol{\Theta})]. \qquad (24)$$

**2.3.2 PINNs for solving Reynolds equation with SS cavitation condition**

The SS cavitation condition requires that the solved pressure field is non-negative according to Eq. (3). We propose two methods to approximate the non-negative and differentiable solution, namely the penalizing SS cavitation condition scheme (denoted as the "Pe-SS" scheme) and the imposing SS cavitation condition scheme (denoted as the "Imp-SS" scheme). The Pe-SS scheme penalizes the residuals of non-negativity by adding a loss term; the Imp-SS scheme constructs a differentiable non-negative output of the neural network, which imposes the condition in PINNs.

Since the condition is constrained to a non-negative pressure field, the pressure converges to zero in the cavitation region for both the value and its gradient. When dealing with the SS cavitation condition, the non-singular term $\frac{\partial H(x^i)}{\partial X}$ at the right end of the Reynolds equation in the cavitation region should be set to zero; otherwise, this term will make the residual at that point non-zero. Before doing this, we must find the cavitation region during the calculation. Since the cavitation region is unknown at the beginning of the calculation, the range of the cavitation region needs to be obtained during the solution process. According to Eq. (3), the criterion $\varepsilon_{P_\Theta}$ for determining the cavitation region is set to the logical value of $(P < 0) \ || \ (P = 0 \ \&\& \ |\nabla P| = 0)$. We develop a formula to determine the cavitation region:
$$\varepsilon_{P_\Theta} = 1 - \varepsilon_+(\boldsymbol{P}) - \varepsilon_-(|\boldsymbol{P}|)\varepsilon_-(|\nabla \boldsymbol{P}|), \qquad (25)$$
where $\varepsilon_+(x)$ and $\varepsilon_-(x)$ are step functions, which are given as
$$\varepsilon_+(x) = \begin{cases} 0, x \geq 0 \\ 1, x < 0 \end{cases}, \varepsilon_-(x) = \begin{cases} 0, x > 0 \\ 1, x \leq 0 \end{cases}. \qquad (26)$$
Thus, the Reynolds equation loss with the SS cavitation condition is expressed as
$$\mathcal{L}_R(\boldsymbol{\Theta}) = \frac{1}{N_{RE}} \sum_{i=1}^{N_{RE}} \left[ \frac{\partial}{\partial X}\left(H^3(x^i)\frac{\partial P_\Theta(x^i)}{\partial X}\right) + \frac{L^2}{B^2}\frac{\partial}{\partial Y}\left(H^3(x^i)\frac{\partial P_\Theta(x^i)}{\partial Y}\right) - \varepsilon_{P_\Theta}\frac{\partial H(x^i)}{\partial X} \right]^2. \qquad (27)$$

- Pe-SS scheme

During the iterative computation, there are always positive and negative pressures, and the final goal is to converge the positive pressure region to the Reynolds equation and set the negative pressure region to zero. The final system of PDEs is satisfied. Then, the non-negativity loss is
$$\mathcal{L}_{SS}(\boldsymbol{\Theta}) = \frac{1}{N_{RE}} \sum_{i=1}^{N_r} \{Relu[-P_\Theta(x^i)]\}^2, \qquad (28)$$
where $Relu$ is the linear rectification function.

- Imp-SS scheme

Inspired by the imposition of the boundary condition scheme [50], we propose Imp-SS to make the PINNs exactly satisfy the non-negativity requirement and construct a differentiable and non-negative function as the activation function of the output layer of the neural network. Here, we use the simple and efficient Relu-square function, which is defined as

$$Relu^2(x) = \begin{cases} x^2, x \geq 0 \\ 0, x < 0 \end{cases}, \tag{29}$$

so that the non-negativity and differential continuity are preserved. Then, the output of the neural network is

$$P_\Theta = \varphi(\tilde{P}^{[M]}) = Relu^2(\tilde{P}^{[M]}). \tag{30}$$

As described above, applying the DB condition is the basis for solving the Reynolds equation. When the two SS schemes and two DB schemes are combined in pairs, we can have four different schemes to simulate the hydrodynamic lubrication with SS cavitation, and the corresponding loss term of each pair can be obtained by summing the two loss terms generated by the two schemes comprising the pair, which are listed in Table 1.

Table 1  Loss terms for solving Reynolds equation with SS cavitation condition in HL-nets

| Scheme | Loss term $\mathcal{L}(\Theta)$ |
| --- | --- |
| Pe-DB & Pe-SS | $[\mathcal{L}_R(\Theta), \mathcal{L}_b(\Theta), \mathcal{L}_{SS}(\Theta)]$ |
| Imp-DB & Pe-SS | $[\mathcal{L}_R(\Theta), \mathcal{L}_{SS}(\Theta)]$ |
| Pe-DB & Imp-SS | $[\mathcal{L}_R(\Theta), \mathcal{L}_b(\Theta)]$ |
| Imp-DB & Imp-SS | $[\mathcal{L}_R(\Theta)]$ |

**2.3.3 PINNs for solving Reynolds equation with JFO cavitation condition**

Compared with the SS cavitation condition, the JFO cavitation condition has an additional FB equation (Eq. (8)), which is used to achieve non-negative pressure. When the JFO cavitation condition is adopted in PINNs, it is natural to apply the algorithm since the cavitation fraction parameter $\theta$ in the FB complementary equation is continuously differentiable. Fig. 2 shows the frame of HL-nets for solving the Reynolds equation with the JFO cavitation condition.

We substitute Eq. (8) into Eq. (9) to define Reynolds equation loss:

$$\mathcal{L}_R(\Theta) = \frac{1}{N_r}\sum_{i=1}^{N_r}\left[\frac{\partial}{\partial X}\left(H^3(x^i)\frac{\partial P_\Theta(x^i)}{\partial X}\right) + \frac{L^2}{B^2}\frac{\partial}{\partial Y}\left(H^3(x^i)\frac{\partial P_\Theta(x^i)}{\partial Y}\right) - \frac{\partial\left((1-\theta_\Theta(x^i))H(x^i)\right)}{\partial X}\right]^2. \tag{31}$$

In addition to the loss term of the Reynolds equation, an additional loss term of the FB complementarity function is added to constrain the complementarity of the pressure and cavitation ratio; the FB equation loss is defined as

$$\mathcal{L}_{FB}(\Theta) = \frac{1}{N_r}\sum_{i=1}^{N_r}\left[P_\Theta(x^i) + \theta_\Theta(x^i) - \sqrt{P_\Theta(x^i)^2 + \theta_\Theta(x^i)^2}\right]^2. \tag{32}$$

Although the FB function (Eq. (9)) contains a constraint for non-negative pressure values, it is not strictly satisfied because it is a soft constraint. When the pressure in the cavitation zone is not exactly equal to zero and fluctuates around zero, the second-order differential term of the pressure of the equation residuals will not be equal to zero, which will lead to a change in the nature of the equation and will significantly impact the cavitation fraction convection equation. To alleviate this problem in PINNs, we can add an additional non-negative constraint of the pressure ($p \geq 0$), like the SS cavitation condition; for implementation in PINNs, the Imp-SS scheme (see Section 2.3.2) can be adopted here due to its good performance (see Section 3.3). With the Imp-SS scheme in the previous section, the Relu-square activation function for the output layer of the pressure will be used to ensure that the pressure is non-negative. Part of the constraint task of the FB function will be satisfied in advance so that the pressure is exactly equal to zero in the cavitation region. Besides, the output layer of the cavitation ratio uses the sigmoid activation function to limit the cavitation ratio $\theta$ to between zero and one.

When Imp-DB is applied, the multiple loss terms are denoted as
$$[\mathcal{L}_i(\Theta)] = [\mathcal{L}_R(\Theta), \mathcal{L}_{FB}(\Theta)]. \tag{33}$$

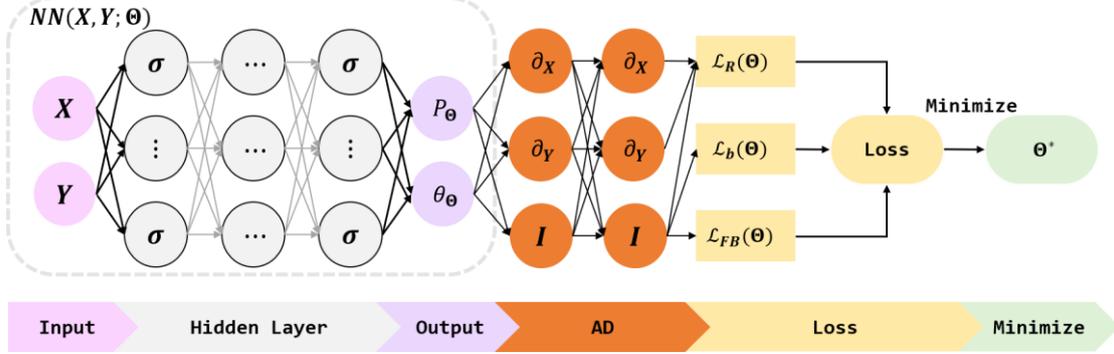

Fig. 2 Frame of HL-nets for solving Reynolds equation with JFO cavitation condition.

## 3. Results and Discussion

To illustrate the prediction performance of HL-nets in solving the Reynolds equation, the flow fields of a bearing are simulated, which is a typical hydrodynamic lubrication problem. We first solve the Reynolds equation without considering cavitation to compare the accuracy of the Pe-DB scheme and Imp-DB scheme for the DB condition; then, for hydrodynamic lubrication involving cavitation, we compare different schemes for SS and JFO cavitation conditions.

## 3.1 Problem setup

In this section, we employ HL-nets to the problem of idealistic oil-lubricated bearing, as shown in Fig. 3, with the dimensions being the same as those of the bearing studied in Ref [60], presented in Table 2. An absolute pressure scale is used, with an ambient pressure of $p_{\partial\Omega} = 72$ kPa and the cavitation pressure arbitrarily taken as $p_{cav} = 0$ kPa. The bearing surface is kept free of inlet holes or grooves for simplicity. In this numerical experiment, the journal is fixed at an eccentricity ratio of 0.6, with the journal rotating at 459 r/min.

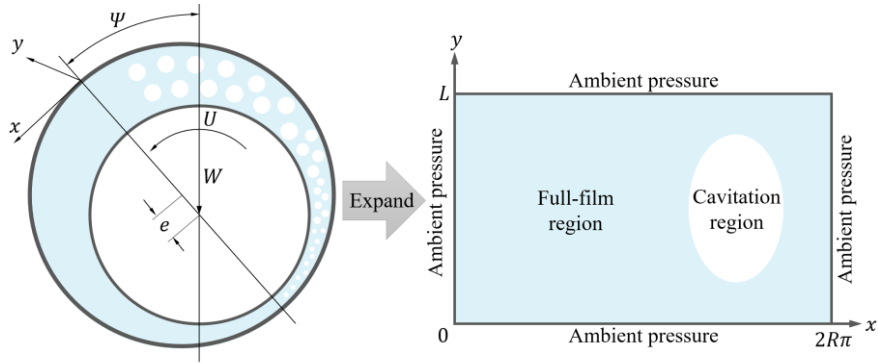

Fig. 3 Geometry of bearing and Reynolds equation domain

Table 2  Structural and operating parameters of bearing

| Parameter | Value |
|---|---|
| Radius $R$ (mm) | 50 |
| Length $L$ (mm) | 8/3R |
| Radius clearance $c$ (m) | 145.5e-6 |
| Viscosity $\eta$ (mP·s) | 0.0127 |
| Eccentricity ratio $e$ | 0.6 |
| Rotating speed $n$ (r/min) | 459 |
| Cavitation pressure $p_{cav}$ (kPa) | 0 |
| Ambient pressure $p_{\partial\Omega}$ (kPa) | 72 |

The film thickness for the journal bearing and DB condition is set to

$$h = e\cos(2\pi(X + 0.5)) + c; h_0 = c, \tag{34}$$

$$P_{\partial\Omega} = \frac{h_0^2}{6\eta UL} p_{\partial\Omega}. \tag{35}$$

The coordinates $\{x, y\}$ are mapped to $\{X, Y\} \in [-0.5, 0.5] \times [-0.5, 0.5]$, because the input parameters of the neural network are more suitable for a symmetric distribution. The computational domain and the dimensionless film thickness are shown in Fig. 4.

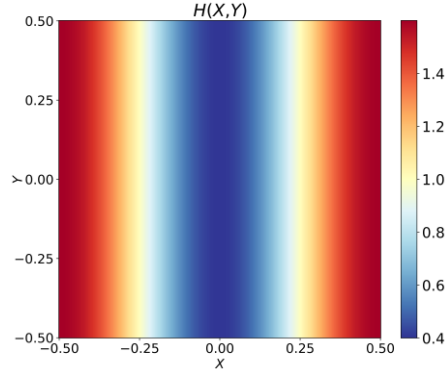

Fig. 4 Computational domain and the dimensionless film thickness.

To evaluate the accuracy of the HL-nets result, the high-precision FEM result is assumed as the reference value. The relative $L_1$ error $L_{1-error}$ and the relative $L_2$ error $L_{2-error}$ are defined as follows:

$$L_{1-error}(P) = \frac{\|\hat{P} - P\|_1}{\|\hat{P}\|_1}, L_{2-error}(P) = \frac{\|\hat{P} - P\|_2}{\|\hat{P}\|_2}, \tag{36}$$

where $\hat{P}$ denotes the pressure inferred by reference, $P$ represents the pressure inferred by PINNs, and $\|\cdot\|_1$ and $\|\cdot\|_2$ are $L_1$-norm and $L_2$-norm, respectively. The absolute error $P_{error}$ is defined as

$$P_{error} = |\hat{P} - P|, \tag{37}$$

and the errors of the cavitation fraction $\theta$ can be calculated in the same way. The reference pressure field $\hat{P}$ and cavitation fraction fields are obtained in this study by discretizing the equations using the high-precision FEM. The SS cavitation condition is applied by setting the negative pressure to zero at each iterative computational step. The JFO cavitation condition is applied using the stabilization method, SUPG [20], and the pressure and cavitation fraction are solved simultaneously using the Fischer-Burmeister-Newton-Schur method [22].

In the following numerical tests, the points selected in the domain and boundary are fixed; a total of 2,500 points in the computational domain are used to compute the equation loss of PINNs, and 100 points on each boundary are used to compute the loss of boundary conditions. All collocation points are generated using uniformly distributed sampling for better stability. The fully connected layer neural network consists of six layers, each containing 20 neurons. In the following case study section, the training procedure is performed by applying the full-batch Adam optimizer with decreasing learning rates $\eta = 10^{-3}$, $10^{-4}$, and $10^{-5}$, and each involves 20,000 epochs.

## 3.2 Comparison of schemes for Dirichlet boundary condition

In this subsection, to verify the treatment of the DB condition, the use of PINNs to solve the Reynolds equation without cavitation is considered, where the unphysical negative pressure is allowed. The DB condition treatment is important because it can greatly affect the accuracy and stability of the calculation with the cavitation in the next section.

In this section, the Pe-DB and Imp-DB schemes are used separately. Since the Pe-DB scheme contains multiple loss terms, the three MTL methods UW, DW, and PCGrad will be armed with the Pe-DB scheme to balance the loss of the Reynolds equation and DB condition. The initial parameter of the UW method is set to $s = [2, -2]$. The DB condition loss $\mathcal{L}_b$ and Reynolds equation loss $\mathcal{L}_R$ of different DB condition schemes and MTL methods during the training process are shown in Fig. 5. From the training graph of losses in Fig. 5, we can observe that the traditional penalizing scheme (Pe-DB) without the MTL method produces both the largest $\mathcal{L}_b$ and the largest $\mathcal{L}_R$, and the $\mathcal{L}_b$ is 1–2 orders of magnitude larger than the results of Pe-DB with the MTL method. Furthermore, the DW method is the most effective in restraining the DB condition loss $\mathcal{L}_b$, while it produces a larger value of equation loss $\mathcal{L}_R$. In terms of equation loss $\mathcal{L}_R$, the Imp-DB scheme and the Pe-DB scheme with the UW method produce the same order of magnitude of the loss, which is smaller than that of Pe-DB with PCGrad or DW. Besides, the Imp-DB scheme has no boundary loss, and its equation loss $\mathcal{L}_R$ can reach a low order of magnitude more quickly during the training process.

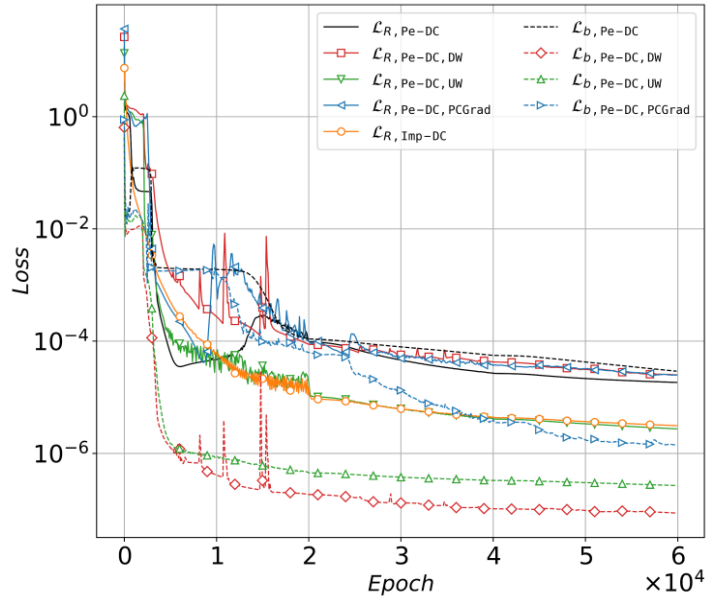

Fig. 5 Training loss of equation residuals $\mathcal{L}_R$ and boundary condition residuals $\mathcal{L}_b$ with different schemes.

Fig. 6 shows the comparison of the pressure value of FEM, the pressure predicted by the Imp-DB scheme, and the pressure predicted by the Pe-DB scheme without an MTL method. It is found that the results of the traditional Pe-DB scheme without an MTL method do not match the FEM results, whereas the results of the Imp-DB scheme agree well with the FEM results. However, when any one of the three MTL methods is employed in the Pe-DB scheme, the accuracy is greatly improved; the results of the Pe-DB scheme with MTL methods are not included in Fig. 6 because the results are indistinguishable from the results of the Imp-DB scheme. Both the Pe-DB scheme with MTL and the Imp-DB scheme produce results with sufficient accuracy for this hydrodynamic lubrication problem without cavitation. In addition, it is difficult to balance the boundary condition loss $\mathcal{L}_b$ and Reynolds equation loss $\mathcal{L}_R$ in this hydrodynamic lubrication problem; so, an MTL method is necessary for the Pe-DB scheme to obtain results with acceptable accuracy.

Fig. 7 shows the error $P_{error}$ between the PINN solution and the solution obtained by FEM. The results of the Imp-DB scheme are the most accurate, and $P_{error}$ at the boundary is equal to zero, which proves that the Imp-DB scheme can ensure that the DB conditions are strictly satisfied. However, for the traditional Pe-DB scheme, the primary error lies in the boundary conditions since the penalizing scheme does not ensure that the boundary conditions are strictly satisfied. Besides, the maximum error on the boundary extends into the bulk area of the computational domain, which indicates that the satisfaction of the boundary conditions affects the global computational accuracy.

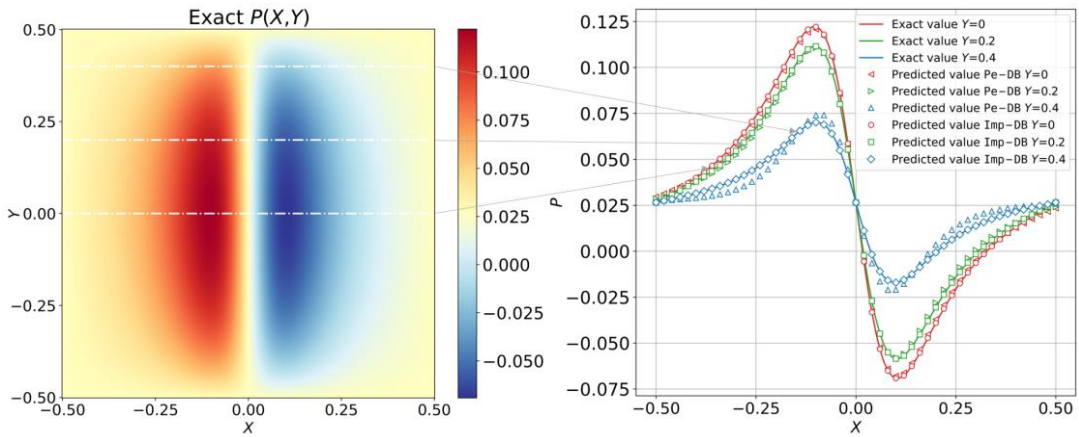

Fig. 6 Comparison of the exact pressure value calculated by FEM, the pressure predicted by the Imp-DB scheme, and the pressure predicted by the traditional Pe-DB scheme. The left part represents the exact pressure field calculated by FEM.

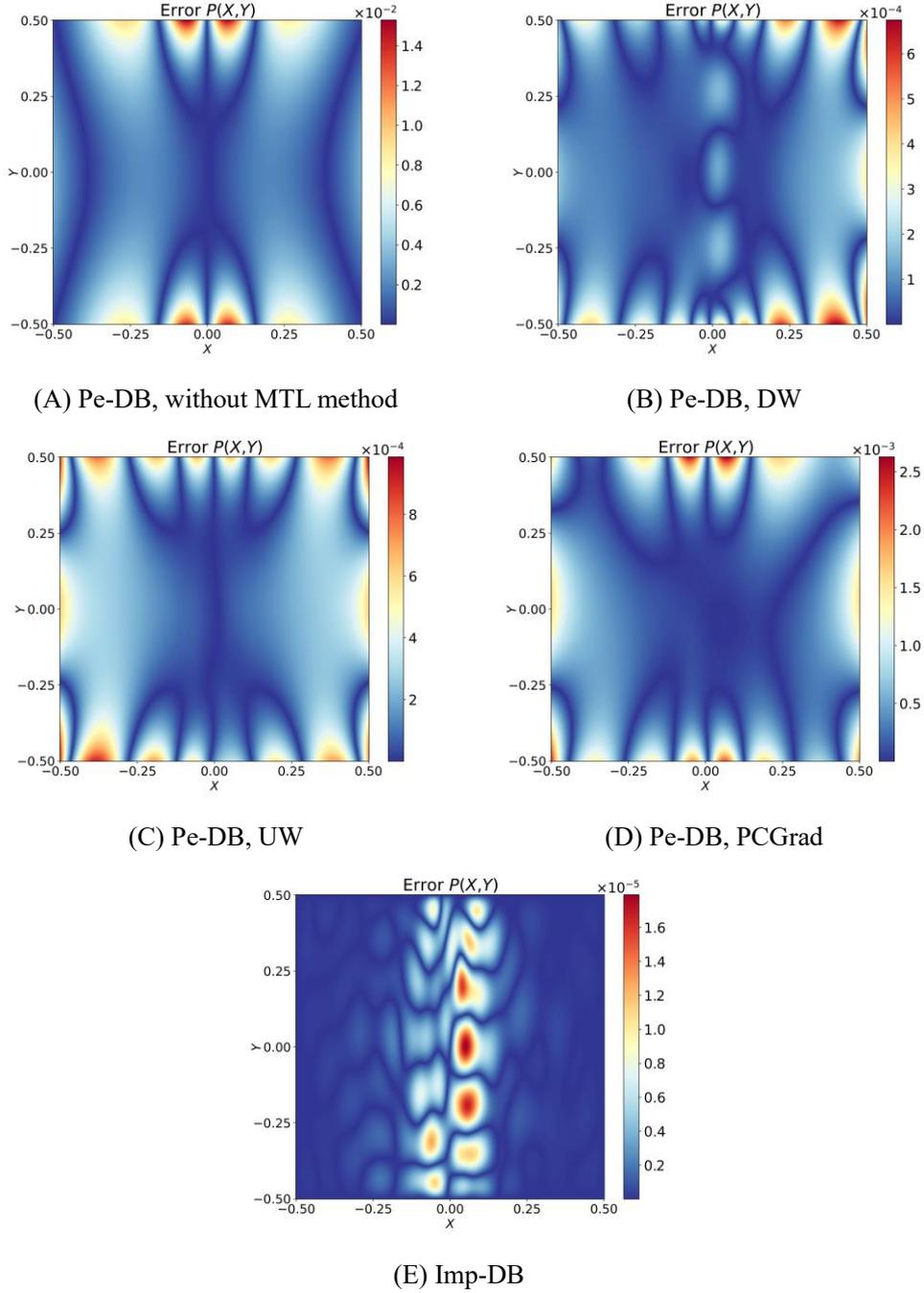

(A) Pe-DB, without MTL method
(B) Pe-DB, DW
(C) Pe-DB, UW
(D) Pe-DB, PCGrad
(E) Imp-DB

Fig. 7 Error $P_{error}$ between the PINN solution and the solution obtained by FEM.

For a more quantitative analysis, the $L_{1-error}$ and $L_{2-error}$ of the pressure field obtained by different schemes are presented in Table 3. The results represent the mean ± standard deviation from five independent runs with independent initial network parameters. According to the results in Table 3, the Pe-DB scheme without an MTL method performs worst, and both its $L_{1-error}$ and $L_{2-error}$ are at least two orders of magnitude larger than those of any other scheme. The three MTL methods adopted in the Pe-DB scheme are listed in descending order of accuracy as DW, UW, and PCGrad in this case. The DW method outperforms the other MTL methods with $L_{1-error} = 2.389\text{e-}03 \pm 6.950\text{e-}04$ and $L_{2-error} = 2.665\text{e-}03 \pm 7.574\text{e-}04$; the relative error of the UW method has the same order of magnitude as that of the DW method, and the relative error of the PCGrad method is an order of magnitude larger than that of other MTL methods. Combined with the training loss curves in Fig. 6, it is found that the boundary loss $\mathcal{L}_b$ are more important than the equation loss $\mathcal{L}_R$. The MTL method that can meet the boundary loss $\mathcal{L}_b$ more precisely can obtain better overall computational accuracy. The DW method prefers to produce a larger weight $\lambda_b$ for boundary loss $\mathcal{L}_b$ than the other two MTL methods; so, it can obtain more precise prediction results. In addition, we consider different sets of initial log variance, and the experimental results show that the log variance $s$ converges to $[-1.186e01 \pm 6.203e-01, -1.706e01 \pm 5.302e-01]$, which indicates that the UW method is not sensitive to the initial weights in HL-nets (see Appendix A). Besides, the PCGrad method can also effectively balance the loss to obtain a result with acceptable accuracy, which is lower than that of other methods. The Imp-DB scheme changes the neural network optimization into single-objective unconstrained optimization, which allows the optimization objective to focus on the loss of the equations without considering the residuals of the output values at the boundary so that the equations can be solved with high accuracy to obtain the pressure field. The $L_{1-error}$ and $L_{2-error}$ of the Imp-DB scheme can reach 3.368e-05 ± 8.274e-06 and 5.508e-05 ± 1.487e-05, respectively, which are two orders of magnitude smaller than those of the traditional Pe-DB scheme with any MTL method. Besides, as shown in Fig. 7(E), the errors of the Imp-DB scheme are mainly concentrated in the region of positive to negative pressure transition in the middle of the flow field, where the pressure gradient is large; so, there are some difficulties for PINNs in solving this region with larger gradients.

Table 3. Performance comparison of Reynolds equation without cavitation

| BC scheme | MTL method | $L_{1-error}$ (P) | $L_{2-error}$ (P) |
|---|---|---|---|
| Pe-DB | - | 1.319e-01 ± 7.362e-02 | 1.489e-01 ± 8.777e-02 |
| Pe-DB | DW | 2.389e-03 ± 6.950e-04 | 2.665e-03 ± 7.574e-04 |
| Pe-DB | UW | 5.322e-03 ± 1.846e-03 | 5.568e-03 ± 1.804e-03 |
| Pe-DB | PCGrad | 9.977e-03 ± 1.347e-03 | 1.154e-02 ± 1.873e-03 |
| **Imp-DB** | **-** | **3.368e-05 ± 8.274e-06** | **5.508e-05 ± 1.487e-05** |

## 3.3 Cavitation problem with Swift-Stieber cavitation condition

The SS cavitation condition is implemented in solving the Reynolds equation in this section, which can make the pressure field closer to physical reality. When applying the SS cavitation condition, the DB condition also should be applied at boundaries. There are two schemes for the DB condition, and two schemes for the SS cavitation condition; by combining these schemes into pairs, we can obtain four different methods for this cavitation problem with the SS cavitation condition: 1) Pe-DB & Pe-SS, 2) Imp-DB & Pe-SS, 3) Pe-DB & Imp-SS, and 4) Imp-DB & Imp-SS. In this section, we examine the computational accuracy of these four different methods with different MTL methods for cavitation problems.

The initial parameter of the UW method is set to $[5, -5, 10]$, $[5, 10]$, and $[5, -5]$ for Pe-DB & Pe-SS, Imp-DB & Pe-SS, and Pe-DB & Imp-SS, respectively. The influence of the initial parameter is discussed in the Appendix A. The loss variation during the PINN training process, as shown in Fig. 8, indicates that the PINNs reached a steady state after $6 \times 10^4$ epochs. Fig. 7 shows the absolute error $P_{error}$ distribution between the PINN solution and the solution obtained by FEM. The relative errors $L_{1-error}$ and $L_{2-error}$ of the pressure field obtained by different schemes are presented in Table 4. The results represent the mean ± standard deviation from five independent runs with independent initial network parameters.

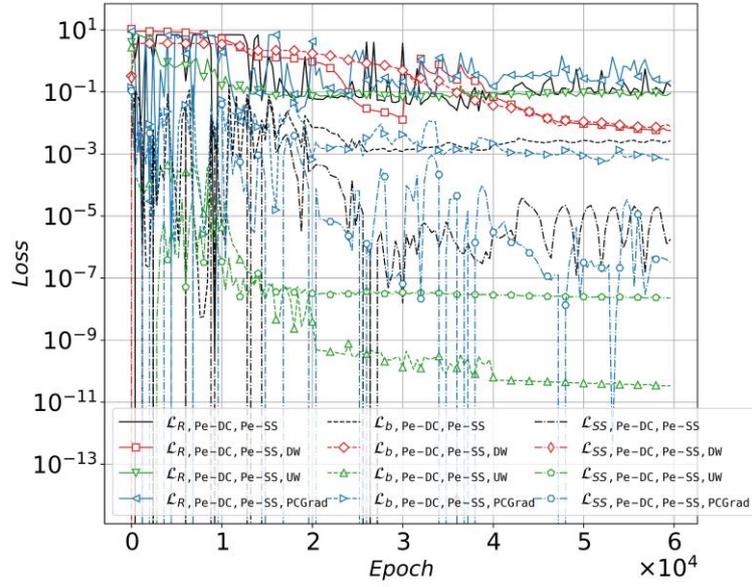

(A) Pe-DB & Pe-SS

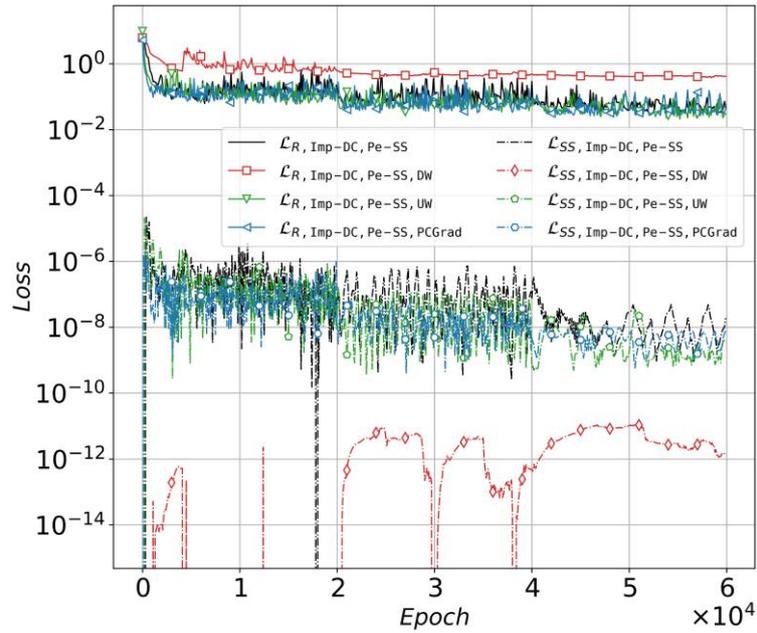

(B) Imp-DB & Pe-SS

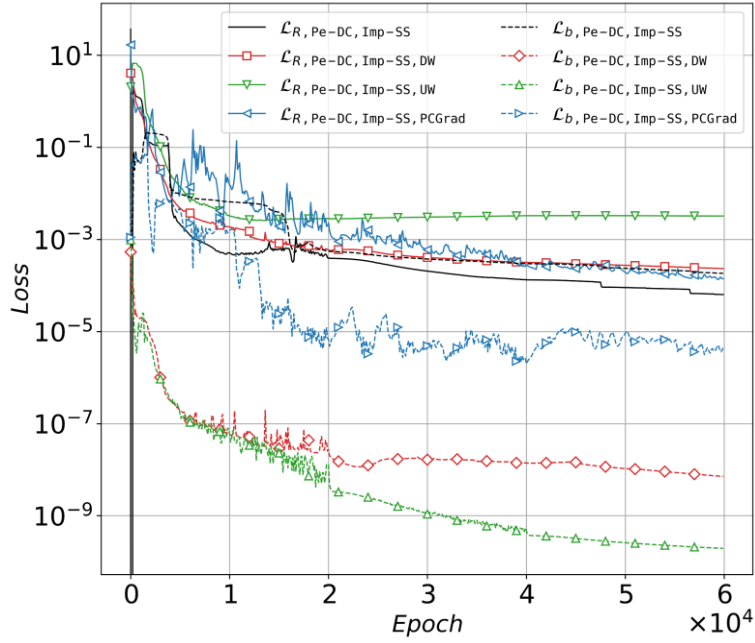

(C) Pe-DB & Imp-SS

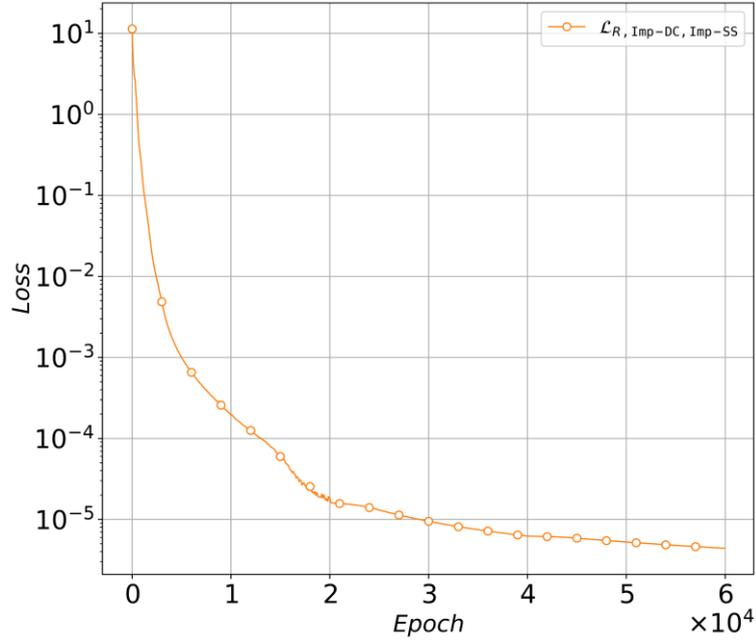

(D) Imp-DB & Imp-SS

Fig. 8 Training losses of different residuals.

The Pe-DB & Pe-SS scheme contains three loss terms, and only combining this scheme with the UW method with the appropriate initial parameters can deliver barely acceptable results, whereas combining this scheme with either of the other two MTL methods results in failure. As shown in Fig. 10(A2), Pe-DB & Pe-SS with UW can better meet the DB condition and ensure that the error is mainly distributed in the transition zone between the pressure extremes and the cavitation zone, which also has a large gradient; the relative $L_1$ error and $L_2$ error of Pe-DB & Pe-SS with the UW scheme are 1.126e-02 ± 5.301e-03 and 1.755e-02 ± 7.533e-03, respectively.

Applying the imposing scheme can reduce the loss terms and thereby simplify optimization.

According to Table 4, the Imp-DB & Pe-SS scheme contains the Reynolds equation loss $\mathcal{L}_R$ and SS loss $\mathcal{L}_{SS}$; the boundary loss $\mathcal{L}_b$ is canceled due to the implementation of the Imp-DB scheme. The Imp-DB & Pe-SS scheme with the UW or PCGrad method can effectively treat the loss of the Reynolds equation and SS cavitation condition. The relative $L_1$ error and the relative $L_2$ error of the UW method can reach 4.697e-03 ± 1.291e-03 and 6.269e-03 ± 1.689e-03, respectively. As shown in Fig. 9(B2), the maximum values of the absolute error $P_{error}$ are located between the peak pressure and the cavitation region, where the pressure obtained by PINNs with the UW method is larger than the reference value. For the Imp-DB & Pe-SS scheme with the PCGrad method, the relative $L_1$ error and the relative $L_2$ error are 6.034e-03 ± 8.113e-04 and 8.704e-03 ± 1.917e-03, respectively. Even though the PCGrad method appears to produce a larger error than the UW method, it does not require any initial parameters, as does UW. From Fig. 9(B3), we can see that the absolute error $P_{error}$ is located between the peak pressure and the cavitation region, and the pressure obtained by PINNs is smaller than the reference value. As discussed in the previous section, the DW method outperforms the other two MTL methods for hydrodynamic lubrication with DB condition, but it produces the worst results when the Pe-SS scheme is adopted. The non-negative constraint cannot be strictly satisfied in the cavitation region when penalizing schemes are used, and we find that the DW method handles the non-negativity loss with too-large weights, thereby preventing the convergence of the loss of the Reynolds equation. As shown in Fig. 8(A, B), the non-negativity loss $\mathcal{L}_{SS}$ of the DW method is far smaller than that of the other two MTL methods, while the equation loss $\mathcal{L}_R$ is far larger than that of the others.

The Pe-DB & Imp-SS scheme with the DW method can obtain high solution accuracy, with relative $L_1$ and $L_2$ errors of 6.999e-04 ± 2.275e-05 and 8.398e-04 ± 2.637e-05, respectively, which are both one order of magnitude smaller than those of the other schemes. As shown in Fig. 9(C1), in addition to the boundary error, the error is also distributed in the transition region between the high-pressure and cavitation regions near the boundary. The DW method is more suitable for dealing with the equation loss and the boundary condition loss than other MTL methods. It can be noted that the PCGrad method achieves a barely acceptable level of accuracy, and the error is mainly distributed in the boundary. With the appropriate initial parameters, the UW method can also achieve high accuracy.

The Pe-DB & Pe-SS scheme can eliminate the boundary and SS cavitation condition residuals, and it obtains the best solution accuracy, with relative $L_1$ and $L_2$ errors of 1.331e-04 ± 5.038e-06 and 3.491e-04 ± 1.240e-04, respectively, which are significantly smaller than those of all the other schemes. For Fig. 9(D), the errors are likewise mainly distributed in the transition part of the high-pressure and cavitation regions near the boundary, where there are fluctuations. This scheme transforms the multi-objective optimization problem into an unconstrained single-objective optimization with the best accuracy and stability. The comparison of the exact pressure and the pressure predicted by the Imp-DB & Imp-SS scheme is shown in Fig. 10, which illustrates the high accuracy of HL-nets for solving the Reynolds equation with the SS cavitation condition.

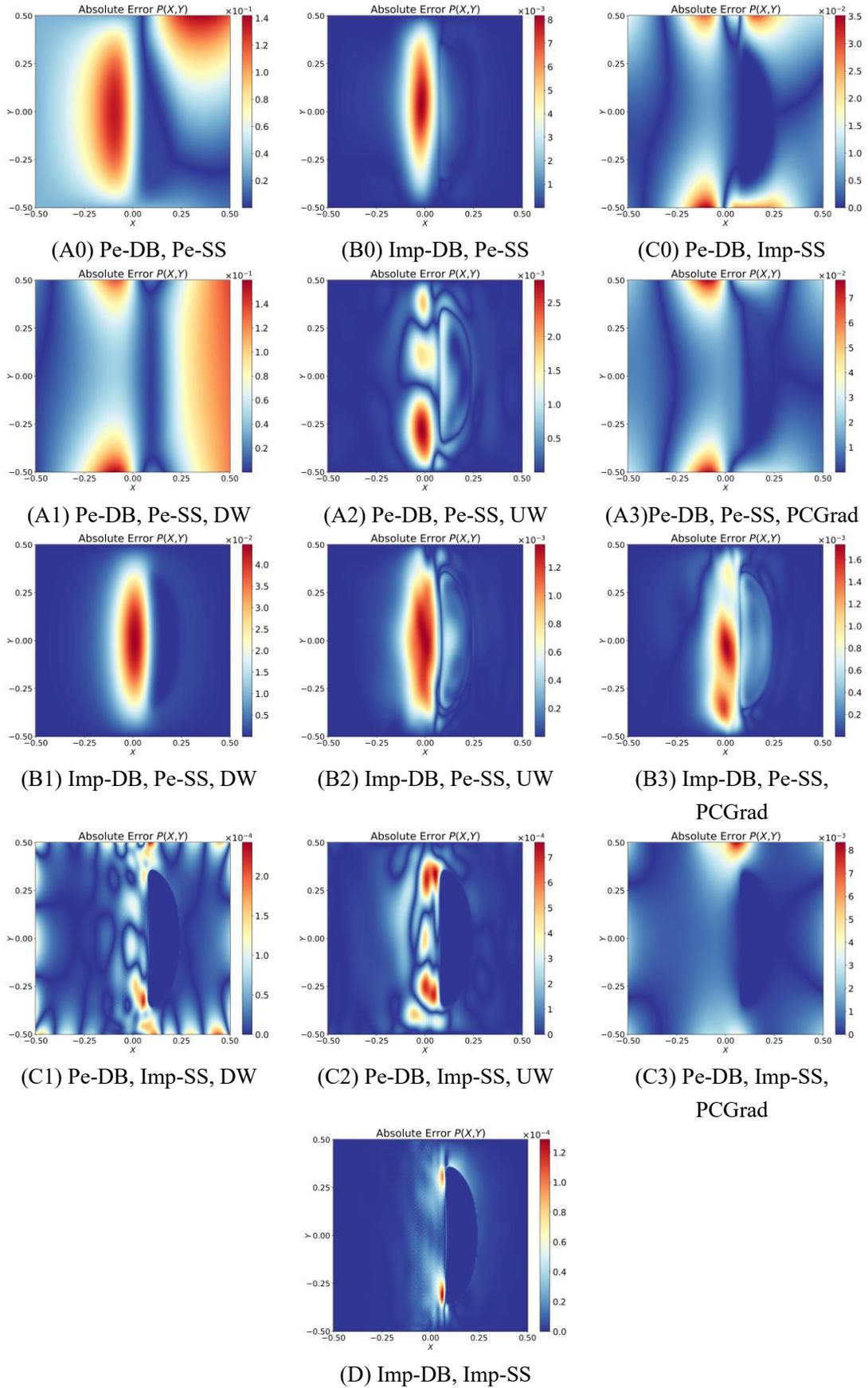

Fig. 9 Absolute error $P_{error}$ between the PINN solution and the solution obtained by FEM.

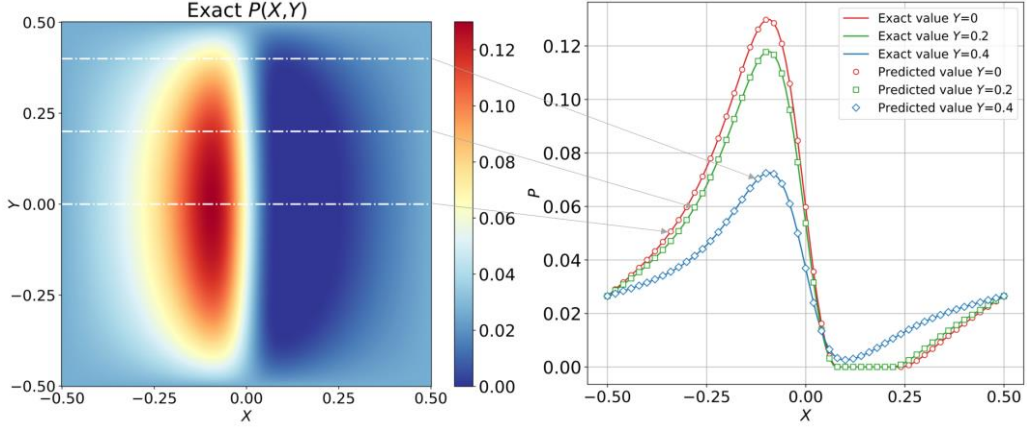

Fig. 10 Comparison of the exact pressure and the pressure predicted by the Pe-DB & Imp-SS scheme. The left part represents the exact pressure field obtained by FEM.

Table 4. Performance comparison of different schemes for the cavitation problem with SS cavitation condition.

| BC scheme | SSC scheme | MTL method | $L_{1-error}(P)$ | $L_{2-error}(P)$ |
|---|---|---|---|---|
| Pe-DB | Pe-SS | - | >1 | >1 |
| Pe-DB | Pe-SS | DW | >1 | >1 |
| Pe-DB | Pe-SS | UW | 1.126e-02 ± 5.301e-03 | 1.755e-02 ± 7.533e-03 |
| Pe-DB | Pe-SS | PCGrad | 3.789e-01 ± 8.934e-02 | 3.877e-01 ± 7.895e-02 |
| Imp-DB | Pe-SS | - | 1.810e-02 ± 1.067e-02 | 3.077e-02 ± 1.856e-02 |
| Imp-DB | Pe-SS | DW | 1.405e-01 ± 3.059e-03 | 2.241e-01 ± 4.758e-03 |
| Imp-DB | Pe-SS | UW | 4.697e-03 ± 1.291e-03 | 6.269e-03 ± 1.689e-03 |
| Imp-DB | Pe-SS | PCGrad | 6.034e-03 ± 8.113e-04 | 8.704e-03 ± 1.917e-03 |
| Pe-DB | Imp-SS | - | 3.806e-01 ± 3.291e-01 | 3.743e-01 ± 3.066e-01 |
| Pe-DB | Imp-SS | DW | 6.999e-04 ± 2.275e-05 | 8.398e-04 ± 2.637e-05 |
| Pe-DB | Imp-SS | UW | 2.007e-03 ± 1.006e-03 | 3.173e-03 ± 1.286e-03 |
| Pe-DB | Imp-SS | PCGrad | 1.467e-02 ± 3.943e-03 | 1.639e-02 ± 4.542e-03 |
| **Imp-DB** | **Imp-SS** | **-** | **1.331e-04 ± 5.038e-06** | **3.491e-04 ± 1.240e-04** |

## 3.4 Cavitation problem with JFO cavitation condition

In this section, the JFO cavitation condition is implemented in solving the Reynolds equation. Our computational experience shows that the Pe-DB scheme fails at solving the pressure and cavitation fraction fields; so, only the Imp-DB scheme is adopted to implement the DB condition in this cavitation problem with the JFO cavitation condition. Besides, the Imp-SS scheme can be used to apply another non-negativity constraint on the pressure in addition to the FB function, which is optional in this section. Since the loss function contains multiple loss terms, the three MTL methods UW, DW, and PCGrad are used to balance the loss. The initial parameter of DW is set to $[5, -3]$. The influence of the initial parameter is discussed in the Appendix A.

Training losses of Reynolds equation residuals $\mathcal{L}_R$ and FB function residuals $\mathcal{L}_{FB}$ with different MTL methods are shown in Fig. 11. The training loss curve is smoother over time when the Imp-SS scheme is applied to strengthen the non-negativity constraint of pressure. In addition, the loss values in Fig. 11(A) are overall smaller than those in Fig. 11(B), which indicates the effectiveness of imposing non-negativity with the JFO cavitation condition. The loss of the Reynolds equation trained by the PCGrad and UW methods are close. The DW method fails to balance the loss of the Reynolds equation and FB function. Too significant a weight factor is assigned to the FB function loss, which makes the complementarity preferentially guaranteed, and the cavitation fraction converges to zero in the whole field, degenerating into the SS cavitation condition of the previous section.

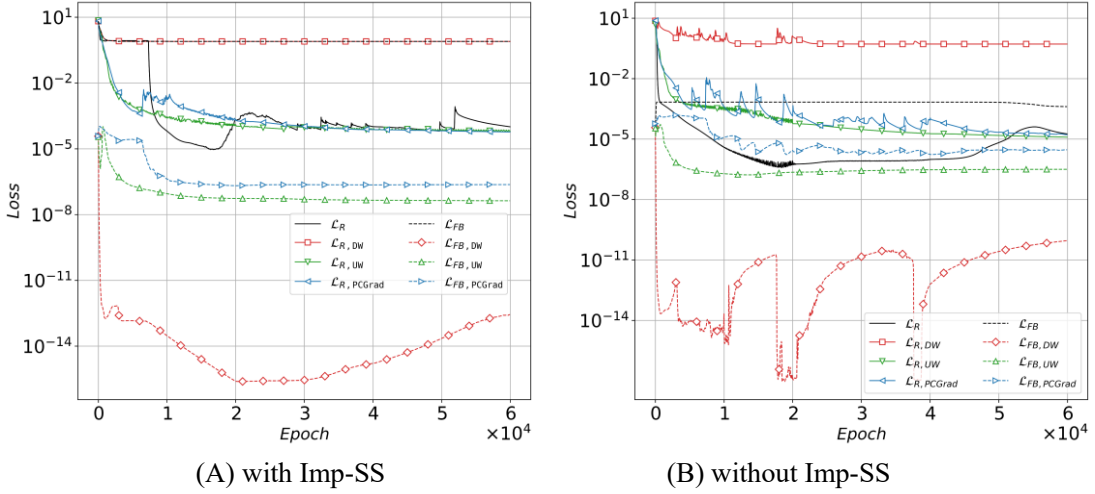

(A) with Imp-SS     (B) without Imp-SS

Fig. 11 Training losses of equation losss $\mathcal{L}_R$ and FB function loss $\mathcal{L}_{FB}$ without or with different MTL methods.

There are evident abrupt gradient extreme regions near the boundary of the cavitation fraction field for this cavitation problem with JFO cavitation condition, and these regions make the cavitation problem challenging for the PINNs to solve. Fig. 12 shows the comparison of the exact value and predicted value with the PCGrad method and non-negativity constraint, and the results of HL-nets agree well with the results of FEM, which proves the accuracy of HL-nets. Fig. 13 shows the absolute error of pressure $P_{error}$ and cavitation fraction $\theta_{error}$ between the HL-nets solution and the solution obtained by FEM. For a more quantitative analysis, the relative errors $L_{1-error}$ and $L_{2-error}$ of the pressure field and cavitation fraction $\theta$ obtained by different schemes are presented in Table 5. The results represent the mean ± standard deviation from five independent runs with independent initial network parameters.

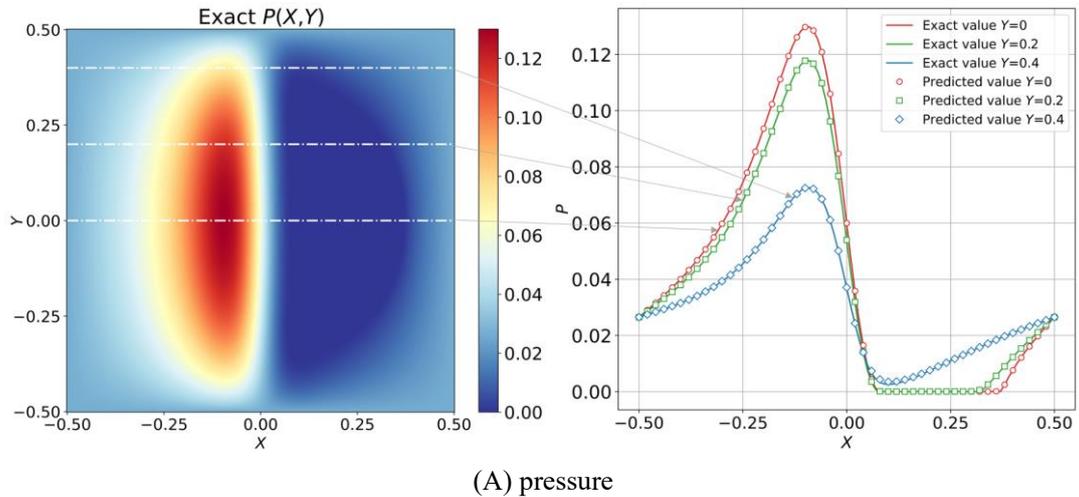

(A) pressure

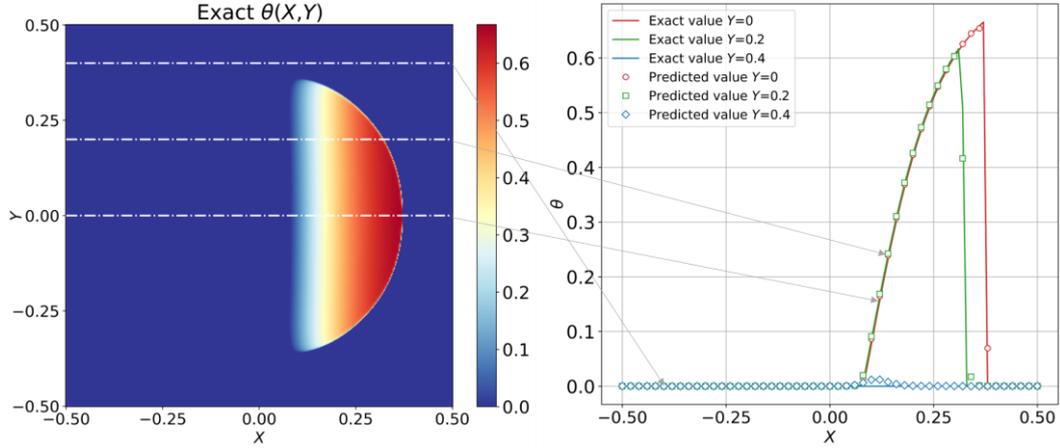

(B) cavitation fraction

Fig. 12 Comparison of the exact value and predicted value with the PCGrad method with non-negativity constraint.

First, according to Table 5 and the error distribution map in Fig. 13, all calculations fail when no MTL method is adopted; so, an MTL method is necessary, regardless of whether or not the Imp-SS scheme is adopted. Besides, the Imp-DB & Imp-SS scheme with the PCGrad method can obtain the best solution; its relative $L_1$ and $L_2$ errors of pressure reach 3.447e-03 ± 1.073e-03 and 5.811e-03 ± 1.816e-03, respectively, and its relative $L_1$ and $L_2$ errors of cavitation fraction $\theta$ reach 8.352e-02 ± 2.366e-02 and 2.657e-02 ± 5.512e-03, respectively. The pressure error and cavitation fraction error are shown in Fig. 13(B2). The error in the cavitation fraction is mainly located in the region of extreme gradient near the boundary of the cavitation region, which makes accurate calculation difficult due to the nonlinear abrupt changes in this region. The corresponding error in the pressure field in this region also remains, except that the main error in the pressure is located in the boundary of the cavitation region near the region of extreme pressure. The Imp-DB & Imp-SS scheme with the UW method can obtain a solution with similar average accuracy as the previous PCGrad method, but the UW method may not be stable without appropriate initial parameters, and the initial parameters of the UW method are further discussed in the Appendix A.

When the non-negativity is not strictly constrained (without the Imp-SS scheme), the left side of the Reynolds equation of Eq. (7) is not equal to zero in the cavitation zone, and it can be seen that the calculation error of the cavitation fraction is larger than that of the Imp-SS scheme because the convective characteristic in the cavitation region is not captured. With the UW method, the relative $L_1$ and $L_2$ errors of pressure reach 3.981e-03 ± 7.242e-04 and 6.173e-03 ± 6.473e-04, respectively, and the relative $L_1$ and $L_2$ errors of cavitation fraction reach 1.393e-01 ± 7.032e-03 and 3.608e-02 ± 1.041e-03, respectively. The pressure and cavitation fraction error distributions are shown in Fig. 13(D1). There are significant errors in pressure and cavitation fraction within and on the boundary of the cavitation region. The PCGrad method achieves a barely acceptable level of accuracy, and its error distribution is similar to that of the UW method.

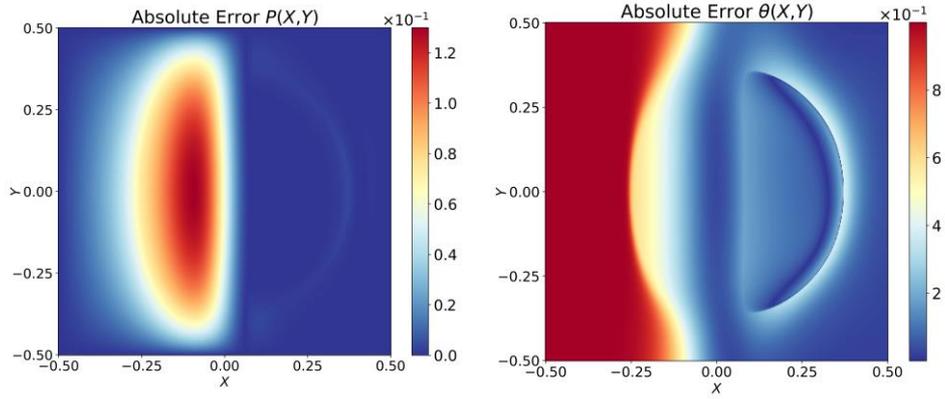
(A0) Without MTL method, Imp-SS

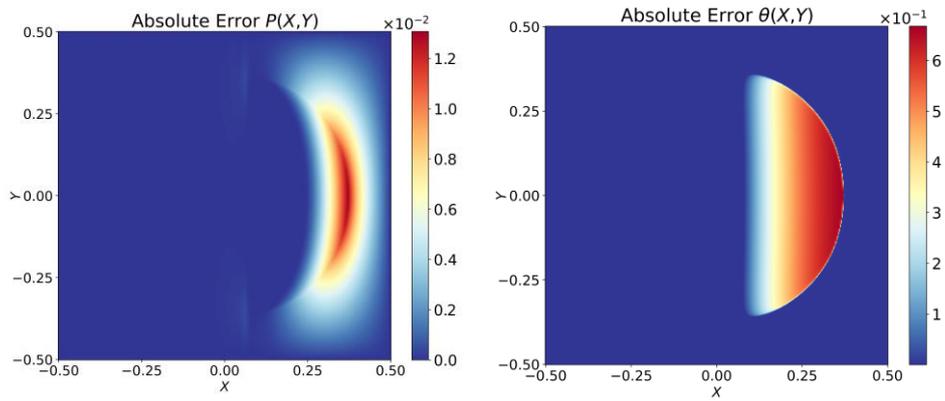
(B0) DW, Imp-SS

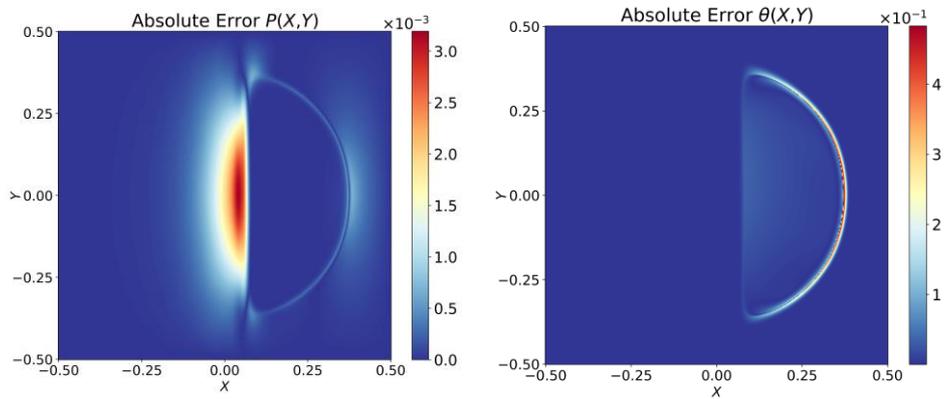
(B1) UW, Imp-SS

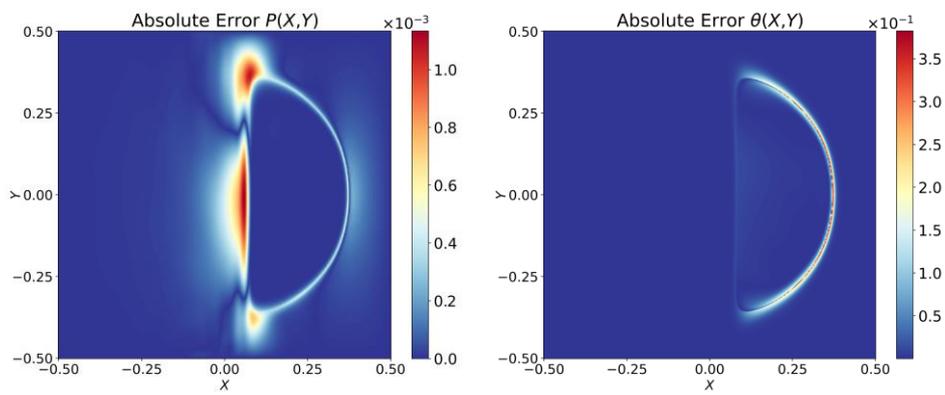
(B2) PCGrad, Imp-SS

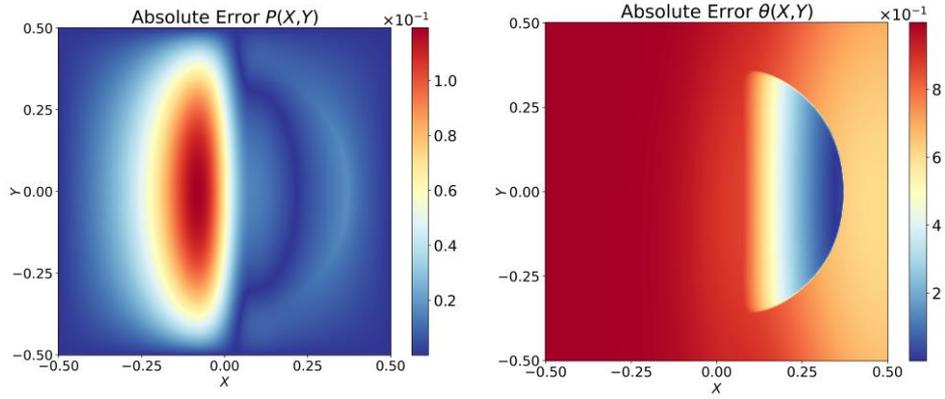
(C) Without MTL method, without Imp-SS

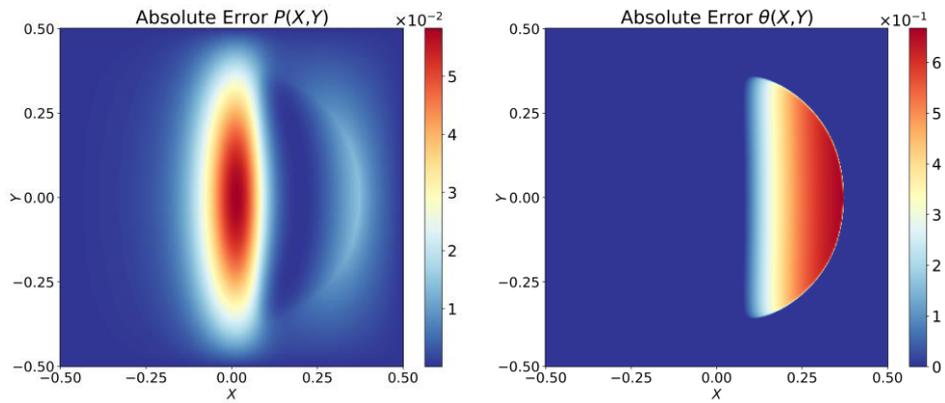
(D0) DW, without Imp-SS

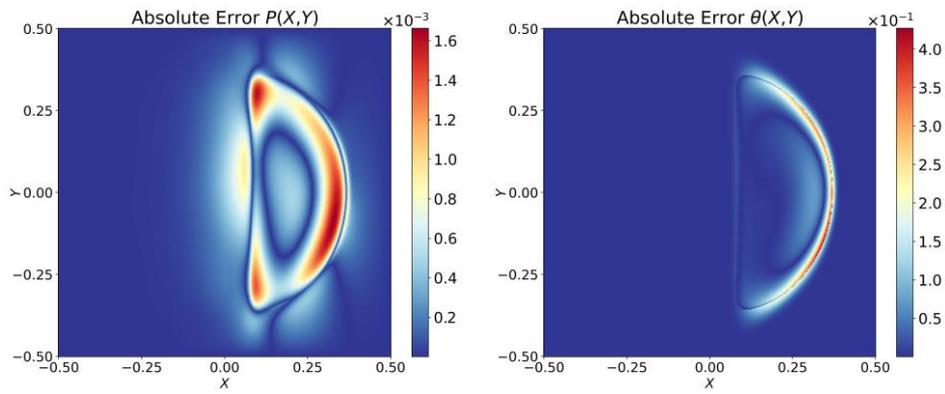
(D1) UW, without Imp-SS

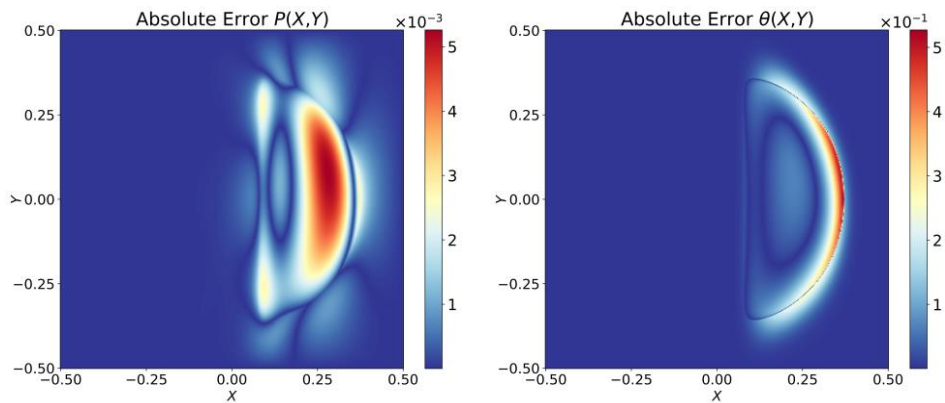
(D2) PCGrad, without Imp-SS

Fig. 13 Absolute error of pressure $P_{error}$ (left) and cavitation fraction $\theta_{error}$ (right) between the HL-nets solution and the solution obtained by FEM.

Table 5 Performance comparison of different schemes in HL-nets for the cavitation problem with JFO cavitation condition.

| Imp-SS | MTL method | $L_{1-error}(P)$ | $L_{2-error}(P)$ | $L_{1-error}(\theta)$ | $L_{1-error}(\theta)$ |
|---|---|---|---|---|---|
| with | - | >0.5 | >0.5 | >1 | >0.5 |
| | DW | 2.780e-02 ± 1.411e-05 | 4.945e-02 ± 9.885e-07 | 1.000 ± 1.103e-06 | 1.816e-01 ± 1.435e-07 |
| | UW | 3.422e-03 ± 3.390e-03 | 5.906e-03 ± 5.608e-03 | 8.686e-02 ± 3.932e-02 | 2.588e-02 ± 7.174e-03 |
| | **PCGrad** | **3.447e-03 ± 1.073e-03** | **5.811e-03 ± 1.816e-03** | **8.352e-02 ± 2.366e-02** | **2.657e-02 ± 5.512e-03** |
| without | - | >0.5 | >0.5 | >1 | >0.5 |
| | DW | 1.513e-01 ± 8.168e-02 | 2.126e-01 ± 1.111e-01 | 8.071e-01 ± 3.859e-01 | 1.479e-01 ± 6.746e-02 |
| | UW | 3.981e-03 ± 7.242e-04 | 6.173e-03 ± 6.473e-04 | 1.393e-01 ± 7.032e-03 | 3.608e-02 ± 1.041e-03 |
| | PCGrad | 1.167e-02 ± 3.845e-04 | 1.753e-02 ± 3.668e-04 | 2.701e-01 ± 6.116e-03 | 5.529e-02 ± 9.951e-04 |

## 4. Conclusions

In this study, we establish a deep learning computational framework, HL-nets, for computing the flow field of hydrodynamic lubrication involving cavitation effects by proposing schemes to apply the SS cavitation condition or the JFO cavitation condition to the PINNs of the Reynolds equation. The results show that HL-nets can highly accurately simulate hydrodynamic lubrication involving cavitation phenomena. The conclusions of this study can be summarized as follows:

(1) For the non-negativity constraint of the SS cavitation condition, the Pe-SS scheme with a loss of the non-negativity and the Imp-SS scheme with a differentiable non-negative function are proposed. By using a differentiable non-negative activation function to constrain the output, the SS cavitation condition can be imposed, and while the loss function contains only one equation loss, the unconstrained optimization process can lead to good computational stability. Both methods have good computational accuracy.

(2) For the complementarity constraint of the JFO cavitation condition, the pressure and cavitation fraction are taken as the neural network outputs, and the residual of the FB equation constrains their complementary relationships. The negativity constraint non-negativity constraint is imposed on the pressure output, partially forcing the FB equation to be satisfied and effectively improving the accuracy of the calculation.

(3) Three MTL methods are applied to balance the newly introduced loss terms described above. The traditional penalizing scheme without the MTL method fails to obtain acceptable results in all cases above, meaning that an appropriate MTL method is needed to improve the accuracy.

# Acknowledge

This work is supported by the National Key R&D Program of China (Grant No. 2020YFB2010000) and the National Natural Science Foundation of China (Grant No. 52275201,12102228).

# Appendix A

In this section, we study the effects of the initial parameters of the UW method in the tests.

(1) Case 1: DB condition: Pe-BC, UW

We test the robustness of the UW method with different initial parameters for the Reynolds equation with the DB condition. The results, presented in Table A1, indicate that the UW method is not sensitive to the initial parameters. However, it is worth noting that the UW method can be solved effectively only when the initial parameters are set so that the initial weight of the DB condition loss is large. When the weight of the boundary loss is not large enough relative to the weight of the equation loss, the iterative calculation of the UW method will intensify the imbalance between the losses. This will cause the weights of the boundary condition loss to be too small, resulting in poor computational accuracy and in the boundary condition not being effectively satisfied.

Table A1. Performance comparison of different initial parameters of the UW method in HL-nets for the Dirichlet boundary condition with the Pe-DB scheme.

| Initial | Final | $L_{1-error}(P)$ | $L_{2-error}(P)$ |
|---|---|---|---|
| [0, 0] | [-1.400e01 ± 2.680e-01, -7.239e00 ± 2.555e-01] | 4.246e-01 ± 7.191e-02 | 4.756e-01±6.546e-02 |
| [1, -1] | [-1.239e01 ± 5.540e-01, -1.491e01 ± 3.859e00] | 8.497e-02 ± 1.652e-01 | 9.484e-02±1.843e-01 |
| [2, -2] | [-1.232e01 ± 5.431e-01, -1.639e01 ± 6.334e-01] | 3.273e-03 ± 1.362e-03 | 3.609e-03±1.356e-03 |
| [3, -3] | [-1.254e01 ± 4.469e-01, -1.669e01 ± 7.285e-01] | 2.998e-03 ± 1.357e-03 | 3.283e-03±1.312e-03 |
| **[4, -4]** | **[-1.161e01 ± 6.657e-01, -1.757e01 ± 4.308e-01]** | **1.550e-03 ± 3.134e-04** | **1.828e-03 ± 3.790e-04** |
| [5, -5] | [-1.096e01 ± 1.639e00, -1.759e01 ± 6.947e-01] | 1.823e-03 ± 3.568e-04 | 2.162e-03 ± 3.893e-04 |

(2) Case 2: Cavitation problem with SS cavitation condition

We test the robustness of the UW method with different initial parameters for the Reynolds equation with the SS cavitation condition. The results are presented in Table A2.

For the Imp-DB & Pe-SS scheme, initializing the log variance so that the weight coefficient corresponding to the non-negativity loss is small leads to more stable and accurate solutions, which indicates that under this condition, the UW method is not sensitive to the initial parameters. The weight coefficient of the non-negativity loss after training iterations is larger than that of the equation loss.

Table A2. Performance comparison of different initial parameters of the UW method in HL-nets for the cavitation problem with SS cavitation condition with Imp-DB & Pe-SS scheme

| Initial | Final | $L_{1-error}(P)$ | $L_{2-error}(P)$ |
|---|---|---|---|
| [5, 4] | [-2.354e00 ± 6.352e-01, -1.817e01 ± 2.717e-04] | 1.889e-01 ± 6.856e-02 | 3.327e-01 ± 1.308e-01 |
| [5, 6] | [-2.508e00 ± 6.113e-01, -1.618e01 ± 6.459e-03] | 1.686e-01 ± 1.418e-01 | 2.954e-01 ± 2.599e-01 |
| [5, 8] | [-2.949e00 ± 3.672e-02, -1.419e01 ± 2.068e-03] | 1.568e-02 ± 1.946e-02 | 2.627e-02 ± 3.453e-02 |
| **[5, 10]** | **[-3.241e00 ± 2.577e-01, -1.219e01 ± 5.315e-04]** | **4.697e-03 ± 1.291e-03** | **6.269e-03 ± 1.689e-03** |
| [5, 12] | [-3.485e00 ± 1.157e-01, -1.020e01 ± 9.091e-05] | 1.906e-02 ± 1.318e-02 | 3.216e-02 ± 2.257e-02 |
| [5, 14] | [-3.189e00 ± 9.527e-02, -8.196e00 ± 1.354e-05] | 7.754e-03 ± 3.689e-03 | 1.163e-02 ± 6.052e-03 |

For the Pe-DB & Imp-SS scheme, there are only equation loss and boundary loss in the loss function; so, the result is similar to that of the Reynolds equation with the DB condition in the previous section. As shown in Table A3, initializing the log variance so that the weight coefficient corresponding to the DB condition loss is small leads to more stable and accurate solutions, which indicates that under this condition, the UW method is not sensitive to the initial parameters. Similarly, the calculation results are poor when the weight coefficients of the boundary condition loss are not large enough relative to the weights of the equation loss.

Table A3. Performance comparison of different initial parameters of the UW method in HL-nets for the cavitation problem with SS cavitation condition with Pe-DB & Imp-SS scheme

| Initial | Final | $L_{1-error}(P)$ | $L_{2-error}(P)$ |
|---|---|---|---|
| [2, -2] | [-1.187e01 ± 1.676e00, -9.621e00 ± 6.277e00] | 7.501e-01 ± 4.329e-01 | 7.501e-01 ± 4.328e-01 |
| **[3, -3]** | **[-8.738e00 ± 6.795e-01, -2.114e01 ± 4.826e-01]** | **4.205e-04 ± 9.926e-05** | **6.625e-04 ± 1.689e-04** |
| [4, -4] | [-8.160e00 ± 2.532e-01, -2.125e01 ± 2.313e-01] | 4.764e-04 ± 7.584e-05 | 7.680e-04 ± 1.257e-04 |
| [5, -5] | [-1.096e01 ± 1.639e00, -1.759e01 ± 6.947e-01] | 9.149e-04 ± 2.693e-04 | 1.627e-03 ± 5.465e-04 |

(3) Case 3: Cavitation problem with JFO cavitation condition

We test the robustness of the UW method with different initial parameters for the Reynolds equation with the JFO cavitation condition. The results are presented in Table A4. With the Imp-SS & Imp-DB scheme, only the Reynolds equation loss and the FB equation loss remain in the loss function. Initializing the log variance so that the weight coefficient corresponding to the FB equation loss is large enough leads to a more stable and accurate solution, which indicates that under this condition, the UW method is not sensitive to the initial parameters. It is noteworthy that the log variance with initial values of [5, −4] is significantly better than that with other initial values. This suggests that fine-tuning the initial parameters in a stable range can be beneficial to the computational results.

Table A4. Performance comparison of different initial parameters of the UW method in HL-nets for the cavitation problem with JFO cavitation condition with Imp-DB & Imp-SS scheme

| Initial | Final | $L_{1-error}(P)$ | $L_{2-error}(P)$ | $L_{1-error}(\theta)$ | $L_{2-error}(\theta)$ |
|---|---|---|---|---|---|
| [5, 0] | [-6.326e00 ± 5.472e00, -1.636e01 ± 3.337e00] | 2.382e-01 ± 2.398e-01 | 3.252e-01 ± 3.200e-01 | 5.560e-01 ± 4.441e-01 | 1.070e-01 ± 7.459e-02 |
| [5, -1] | [-7.952e00 ± 4.131e00, -1.557e01 ± 2.749e00] | 1.996e-01 ± 3.420e-01 | 2.319e-01 ± 3.952e-01 | 8.369e-01 ± 1.327e00 | 1.113e-01 ± 1.488e-01 |
| [5, -2] | [-1.014e01 ± 8.123e-01, -1.767e01 ± 5.577e-01] | 3.355e-03 ± 2.536e-03 | 5.922e-03 ± 4.537e-03 | 7.881e-02 ± 3.575e-02 | 2.292e-02 ± 7.518e-03 |
| [5, -3] | [-1.067e01 ± 3.462e-01, -1.772e01 ± 1.671e-01] | 3.422e-03 ± 3.390e-03 | 5.906e-03 ± 5.608e-03 | 8.686e-02 ± 3.932e-02 | 2.588e-02 ± 7.174e-03 |
| **[5, -4]** | **[-1.001e01 ± 3.811e-02, -1.854e01 ± 2.435e-02]** | **8.103e-04 ± 4.805e-04** | **1.193e-03 ± 6.178e-04** | **5.381e-02 ± 1.400e-02** | **2.032e-02 ± 3.371e-03** |
| [5, -5] | [-9.098e00 ± 4.288e-01, -1.793e01 ± 3.841e-01] | 3.171e-03 ± 2.729e-03 | 5.321e-03 ± 4.431e-03 | 8.133e-02 ± 2.730e-02 | 2.301e-02 ± 5.909e-03 |

# Appendix B

The dimensionless total load capacity $W$ and the attitude angle $\Psi$ are the critical performance parameters of the bearing, which gives:

$$\begin{bmatrix} W\cos\Psi \\ W\sin\Psi \end{bmatrix} = \int_{-\frac{1}{2}}^{\frac{1}{2}} \int_{-\frac{1}{2}}^{\frac{1}{2}} P \begin{bmatrix} \cos(2\pi X) \\ \sin(2\pi X) \end{bmatrix} dXdY \quad (B.1)$$

The relative absolute error $L_1$ is defined as follows:

$$L_1(W) = \frac{|\widehat{W} - W|}{|\widehat{W}|} \quad (B.2)$$

where $\widehat{W}$ denotes the dimensionless total load capacity inferred by reference, $W$ represents the the dimensionless total load capacity inferred by PINNs. The relative absolute error of the maximum pressure $L_1(P_{max})$, the relative absolute error of the maximum cavitation fraction $L_1(\theta_{max})$ and the relative absolute error of the attitude angle $L_1(\Psi)$ are defined similarly to the relative absolute error of the dimensionless total load capacity. To test the bearing performance obtained by HL-net for the cavitation problem with JFO cavitation condition, the relative absolute error $L_1$ are presented in Table B1. The results represent the mean ± standard deviation from five independent runs with independent initial network parameters.

Table B1 Performance parameters comparison of different schemes in HL-nets for the cavitation problem with JFO cavitation condition.

| Imp-SS | MTL method | $L_1(P_{max})$ | $L_1(\theta_{max})$ | $L_1(W)$ | $L_1(\Psi)$ |
|---|---|---|---|---|---|
| | - | >0.5 | >0.5 | >0.5 | >1 |
| with | DW | 9.966e-04 ± 1.951e-05 | 1.000e00 | 4.655e-02 ± 4.112e-05 | 9.728e-03 ± 2.752e-05 |
| | UW | 1.455e-03 ± 1.347e-03 | 2.339E-02 ± 1.182E-02 | 1.232e-03 ± 8.940e-04 | 3.068e-03 ± 3.505e-03 |
|  | PCGrad | 1.974e-03 ± 9.759e-04 | 3.006E-02 ± 2.548E-02 | 1.165e-03 ± 3.595e-04 | 3.821e-03 ± 2.360e-03 |
| without | - | >0.5 | >0.5 | >0.5 | >1 |
|  | DW | 9.670e-02 ± 9.421e-02 | 8.121e-01 ± 3.759e-01 | 7.617e-02 ± 4.958e-02 | 1.472e-01 ± 8.639e-02 |
|  | UW | 1.427e-03 ± 3.085e-04 | 3.345e-02 ± 1.113e-02 | 3.327e-03 ± 6.206e-04 | 4.951e-04 ± 3.795e-04 |
|  | PCGrad | 2.623e-03 ± 3.364e-04 | 4.497e-02 ± 2.331e-02 | 1.144e-02 ± 9.904e-04 | 2.092e-03 ± 2.988e-04 |

| Nomenclature | | | |
|---|---|---|---|
| PINNs | physics-informed neural networks | PDEs | partial differential equations |
| HL-nets | physics-informed neural networks for hydrodynamic lubrication with cavitation | SS | Swift-Stieber |
| JFO | Jakobsson-Floberg-Olsson | FB | Fischer-Burmeister |
| MTL | multi-task learning | DB | Dirichlet boundary |
| $\rho$ | density of the fluid | $p$ | pressure |
| $h$ | film thickness | $\mu$ | viscosity of the fluid |
| $U$ | relative sliding velocity | $p_{\partial\Omega}$ | ambient pressure |
| $p_{cav}$ | cavitation pressure | $\theta$ | cavitation fraction |
| $\rho_0$ | constant density in the full-film region | $L$ | length of the lubrication region |
| $B$ | width of the lubrication region | $h_0$ | minimum film thickness |
| $R$ | radius of bearings | $x, y$ | coordinate |
| $X, Y$ | dimensionless coordinate | SUPG | Streamline Upwind/Petrov-Galerkin |
| FEM | finite element method | $W^{[m]}$ | trainable weights at the $m$-th layer |
| $b^{[m]}$ | trainable biases at the $m$-th layer | $\Theta$ | trainable parameter set of the neural network |
| $\sigma$ | activation function | $\mathcal{L}_R$ | Reynolds equation loss |
| $N_R$ | number of data points for the bulk domain | $\lambda_R, \lambda_i$ | weight parameters |
| $\eta$ | learning rate | $\mathcal{L}$ | total loss |
| $\varphi$ | function used to apply the constraint condition | DW | dynamic weight |
| UW | uncertainty weight | PCGrad | projecting conflicting gradient |
| $\alpha$ | hyperparameter in DW | $N_T$ | number of loss terms |
| $s, s_i$ | log variance | Pe-DB | penalizing scheme for the DB condition |
| Imp-DB | imposing scheme for the DB condition | $\mathcal{L}_b$ | DB condition loss |
| $\phi$ | distance function with a value of zero at boundaries | $P_{\partial\Omega}$ | boundary value function |
| Pe-SS | penalizing SS cavitation condition | Imp-SS | imposing SS cavitation condition |
| $\varepsilon_+, \varepsilon_-$ | step functions | $\varepsilon_{P_\Theta}$ | formula to determine the cavitation region |
| $\mathcal{L}_{SS}$ | non-negativity loss | $\mathcal{L}_{FB}$ | the FB equation loss |
| $c$ | radius clearance | $e$ | eccentricity ratio |
| $n$ | rotating speed | $L_{1-error}, L_{2-error}$ | the relative $L_1$ error and the relative $L_2$ error |
| $P_{error}$ | absolute error | | |

# Reference

[1] Gropper D, Wang L, Harvey TJ. Hydrodynamic lubrication of textured surfaces: A review of modeling techniques and key findings. Tribol Int 2016;94:509–29. https://doi.org/10.1016/j.triboint.2015.10.009.

[2] IV. On the theory of lubrication and its application to Mr. Beauchamp tower's experiments, including an experimental determination of the viscosity of olive oil. Philos Trans R Soc Lond 1886;177:157–234. https://doi.org/10.1098/rstl.1886.0005.

[3] Szeri AZ. Fluid film lubrication. Cambridge university press; 2010.

[4] Hamrock BJ, Schmid SR, Jacobson BO. Fundamentals of fluid film lubrication. CRC press; 2004.

[5] Braun MJ, Hannon WM. Cavitation formation and modelling for fluid film bearings: A review. Proc Inst Mech Eng Part J J Eng Tribol 2010;224:839–63. https://doi.org/10.1243/13506501JET772.

[6] Xing C, Braun MJ, Li H. A Three-Dimensional Navier-Stokes–Based Numerical Model for Squeeze-Film Dampers. Part 1—Effects of Gaseous Cavitation on Pressure Distribution and Damping Coefficients without Consideration of Inertia. Tribol Trans 2009;52:680–94. https://doi.org/10.1080/10402000902913303.

[7] Swift HW. THE STABILITY OF LUBRICATING FILMS IN JOURNAL BEARINGS. (INCLUDES APPENDIX). Minutes Proc Inst Civ Eng 1932;233:267–88. https://doi.org/10.1680/imotp.1932.13239.

[8] Körner K. Dr.-Ing. Wilhelm Stieber, Das Schwimmlager. Hydrodynamische Theorie des Gleitlagers. VII + 106 S. m. 12 Zahlent. u. 42 Abb. Berlin 1933, VDI-Verlag. Preis 6 M, VDI-Mitgl. 5,50 M. ZAMM - Z Für Angew Math Mech 1933;13:391–391. https://doi.org/10.1002/zamm.19330130521.

[9] Gustafsson T, Rajagopal KR, Stenberg R, Videman J. An adaptive finite element method for the inequality-constrained Reynolds equation. Comput Methods Appl Mech Eng 2018;336:156–70. https://doi.org/10.1016/j.cma.2018.03.004.

[10] Olsson K. Cavitation in Dynamically loaded Bearings, 1965.

[11] Jakobsson B, Floberg L. The finite journal bearing, considering vaporization (Das Gleitlager von endlicher Breite mit Verdampfung), 1957.

[12] Elrod HG. A Cavitation Algorithm. J Lubr Technol 1981;103:350–4. https://doi.org/10.1115/1.3251669.

[13] Miraskari M, Hemmati F, Jalali A, Alqaradawi MY, Gadala MS. A Robust Modification to the Universal Cavitation Algorithm in Journal Bearings. J Tribol 2017;139:031703. https://doi.org/10.1115/1.4034244.

[14] Vijayaraghavan D, Keith TG. Development and Evaluation of a Cavitation Algorithm. Tribol Trans 1989;32:225–33. https://doi.org/10.1080/10402008908981882.

[15] Bonneau D, Huitric J, Tournerie B. Finite Element Analysis of Grooved Gas Thrust Bearings and Grooved Gas Face Seals. J Tribol 1993;115:348–54. https://doi.org/10.1115/1.2921642.

[16] Hajjam M, Bonneau D. A transient finite element cavitation algorithm with application to radial lip seals. Tribol Int 2007;40:1258–69. https://doi.org/10.1016/j.triboint.2007.01.018.

[17] Fesanghary M, Khonsari MM. A Modification of the Switch Function in the Elrod Cavitation Algorithm. J Tribol 2011;133:024501. https://doi.org/10.1115/1.4003484.

[18] Ausas RF, Jai M, Buscaglia GC. A Mass-Conserving Algorithm for Dynamical Lubrication Problems With Cavitation. J Tribol 2009;131:031702. https://doi.org/10.1115/1.3142903.

[19] Ransegnola T, Sadeghi F, Vacca A. An Efficient Cavitation Model for Compressible Fluid Film Bearings. Tribol Trans 2021;64:434–53. https://doi.org/10.1080/10402004.2020.1853864.

[20] Ausas R, Ragot P, Leiva J, Jai M, Bayada G, Buscaglia GC. The Impact of the Cavitation Model in the Analysis of Microtextured Lubricated Journal Bearings. J Tribol 2007;129:868–75. https://doi.org/10.1115/1.2768088.

[21] Sobhi S, Nabhani M, Zarbane K, El Khlifi M. Cavitation in oscillatory porous squeeze film: a numerical approach. Ind Lub Tribol 2022;74:636–44. https://doi.org/10.1108/ILT-09-2021-0376.

[22] Meng X, Bai S, Peng X. An efficient adaptive finite element method algorithm with mass conservation for analysis of liquid face seals. J Zhejiang Univ Sci A 2014;15:172–84. https://doi.org/10.1631/jzus.A1300328.

[23] Giacopini M, Fowell MT, Dini D, Strozzi A. A Mass-Conserving Complementarity Formulation to Study Lubricant Films in the Presence of Cavitation. J Tribol 2010;132:041702. https://doi.org/10.1115/1.4002215.

[24] Woloszynski T, Podsiadlo P, Stachowiak GW. Efficient Solution to the Cavitation Problem in Hydrodynamic Lubrication. Tribol Lett 2015;58:18. https://doi.org/10.1007/s11249-015-0487-4.

[25] Silva A, Lenzi V, Cavaleiro A, Carvalho S, Marques L. FELINE: Finite element solver for hydrodynamic lubrication problems using the inexact Newton method. Comput Phys Commun 2022;279:108440. https://doi.org/10.1016/j.cpc.2022.108440.

[26] Biboulet N, Lubrecht AA. Efficient solver implementation for reynolds equation with mass-conserving cavitation. Tribol Int 2018;118:295–300. https://doi.org/10.1016/j.triboint.2017.10.008.

[27] Geng Y, Chen W. Multiscale method of modelling surface texture with mass-conserving cavitation model. Tribol Int 2022;173:107663. https://doi.org/10.1016/j.triboint.2022.107663.

[28] Raissi M, Perdikaris P, Karniadakis GE. Physics-informed neural networks: A deep learning

framework for solving forward and inverse problems involving nonlinear partial differential equations. J Comput Phys 2019;378:686–707. https://doi.org/10.1016/j.jcp.2018.10.045.

[29] Krishnapriyan AS, Gholami A. Characterizing possible failure modes in physics-informed neural networks n.d.:13.

[30] Chiu P-H, Wong JC, Ooi C, Dao MH, Ong Y-S. CAN-PINN: A fast physics-informed neural network based on coupled-automatic–numerical differentiation method. Comput Methods Appl Mech Eng 2022;395:114909. https://doi.org/10.1016/j.cma.2022.114909.

[31] Bai J, Zhou Y, Ma Y, Jeong H, Zhan H, Rathnayaka C, et al. A general Neural Particle Method for hydrodynamics modeling. Comput Methods Appl Mech Eng 2022;393:114740. https://doi.org/10.1016/j.cma.2022.114740.

[32] Sun L, Gao H, Pan S, Wang J-X. Surrogate modeling for fluid flows based on physics-constrained deep learning without simulation data. Comput Methods Appl Mech Eng 2020;361:112732. https://doi.org/10.1016/j.cma.2019.112732.

[33] Ranade R, Hill C, Pathak J. DiscretizationNet: A machine-learning based solver for Navier–Stokes equations using finite volume discretization. Comput Methods Appl Mech Eng 2021;378:113722. https://doi.org/10.1016/j.cma.2021.113722.

[34] Buhendwa AB, Adami S, Adams NA. Inferring incompressible two-phase flow fields from the interface motion using physics-informed neural networks. Mach Learn Appl 2021;4:100029. https://doi.org/10.1016/j.mlwa.2021.100029.

[35] Wessels H, Weißenfels C, Wriggers P. The neural particle method – An updated Lagrangian physics informed neural network for computational fluid dynamics. Comput Methods Appl Mech Eng 2020;368:113127. https://doi.org/10.1016/j.cma.2020.113127.

[36] Guo H, Zhuang X, Rabczuk T. A Deep Collocation Method for the Bending Analysis of Kirchhoff Plate. Comput Mater Contin 2019;59:433–56. https://doi.org/10.32604/cmc.2019.06660.

[37] Haghighat E, Raissi M, Moure A, Gomez H, Juanes R. A physics-informed deep learning framework for inversion and surrogate modeling in solid mechanics. Comput Methods Appl Mech Eng 2021;379:113741. https://doi.org/10.1016/j.cma.2021.113741.

[38] Cai S, Wang Z, Wang S, Perdikaris P, Karniadakis GE. Physics-Informed Neural Networks for Heat Transfer Problems. J Heat Transf 2021;143:060801. https://doi.org/10.1115/1.4050542.

[39] Zhao X, Shirvan K, Salko RK, Guo F. On the prediction of critical heat flux using a physics-informed machine learning-aided framework. Appl Therm Eng 2020;164:114540. https://doi.org/10.1016/j.applthermaleng.2019.114540.

[40] Kadeethum T, Jørgensen TM, Nick HM. Physics-informed neural networks for solving nonlinear diffusivity and Biot's equations. PLOS ONE 2020;15:e0232683. https://doi.org/10.1371/journal.pone.0232683.

[41] Hanna JM, Aguado JV, Comas-Cardona S, Askri R, Borzacchiello D. Residual-based adaptivity for two-phase flow simulation in porous media using Physics-informed Neural Networks. Comput Methods Appl Mech Eng 2022;396:115100. https://doi.org/10.1016/j.cma.2022.115100.

[42] Chen F, Sondak D, Protopapas P, Mattheakis M, Liu S, Agarwal D, et al. NeuroDiffEq: A Python package for solving differential equations with neural networks. J Open Source Softw 2020;5:1931. https://doi.org/10.21105/joss.01931.

[43] Lu L, Meng X, Mao Z, Karniadakis GE. DeepXDE: A Deep Learning Library for Solving Differential Equations. SIAM Rev 2021;63:208–28. https://doi.org/10.1137/19M1274067.

[44] McClenny LD, Haile MA, Braga-Neto UM. TensorDiffEq: Scalable Multi-GPU Forward and Inverse Solvers for Physics Informed Neural Networks 2021.

[45] Almqvist A. Fundamentals of Physics-Informed Neural Networks Applied to Solve the Reynolds Boundary Value Problem. Lubricants 2021;9:82. https://doi.org/10.3390/lubricants9080082.

[46] Zhao Y, Guo L, Wong PPL. Application of physics-informed neural network in the analysis of hydrodynamic lubrication. Friction 2022. https://doi.org/10.1007/s40544-022-0658-x.

[47] Li L, Li Y, Du Q, Liu T, Xie Y. ReF-nets: Physics-informed neural network for Reynolds equation of gas bearing. Comput Methods Appl Mech Eng 2022;391:114524. https://doi.org/10.1016/j.cma.2021.114524.

[48] Abadi M, Barham P, Chen J, Chen Z, Davis A, Dean J, et al. TensorFlow: A system for large-scale machine learning n.d.:21.

[49] Paszke A, Gross S, Massa F, Lerer A, Bradbury J, Chanan G, et al. PyTorch: An Imperative Style, High-Performance Deep Learning Library n.d.:12.

[50] Sukumar N, Srivastava A. Exact imposition of boundary conditions with distance functions in physics-informed deep neural networks. Comput Methods Appl Mech Eng 2022;389:114333. https://doi.org/10.1016/j.cma.2021.114333.

[51] Dong S, Ni N. A method for representing periodic functions and enforcing exactly periodic boundary conditions with deep neural networks. J Comput Phys 2021;435:110242. https://doi.org/10.1016/j.jcp.2021.110242.

[52] Wang S, Yu X, Perdikaris P. When and why PINNs fail to train: A neural tangent kernel perspective. J Comput Phys 2022;449:110768. https://doi.org/10.1016/j.jcp.2021.110768.

[53] Wang S, Teng Y, Perdikaris P. Understanding and Mitigating Gradient Flow Pathologies in Physics-Informed Neural Networks. SIAM J Sci Comput 2021;43:A3055–81. https://doi.org/10.1137/20M1318043.

[54] Vandenhende S, Georgoulis S, Van Gansbeke W, Proesmans M, Dai D, Van Gool L. Multi-Task Learning for Dense Prediction Tasks: A Survey. IEEE Trans Pattern Anal Mach Intell 2021:1–1. https://doi.org/10.1109/TPAMI.2021.3054719.

[55] Jin X, Cai S, Li H, Karniadakis GE. NSFnets (Navier-Stokes flow nets): Physics-informed neural networks for the incompressible Navier-Stokes equations. J Comput Phys 2021;426:109951. https://doi.org/10.1016/j.jcp.2020.109951.

[56] Cipolla R, Gal Y, Kendall A. Multi-task Learning Using Uncertainty to Weigh Losses for Scene Geometry and Semantics. 2018 IEEECVF Conf. Comput. Vis. Pattern Recognit., Salt Lake City, UT, USA: IEEE; 2018, p. 7482–91. https://doi.org/10.1109/CVPR.2018.00781.

[57] Xiang Z, Peng W, Liu X, Yao W. Self-adaptive loss balanced Physics-informed neural networks. Neurocomputing 2022;496:11–34. https://doi.org/10.1016/j.neucom.2022.05.015.

[58] Yu T, Kumar S, Gupta A, Levine S, Hausman K, Finn C. Gradient Surgery for Multi-Task Learning 2020.

[59] Thanasutives P, Numao M, Fukui K. Adversarial Multi-task Learning Enhanced Physics-informed Neural Networks for Solving Partial Differential Equations. 2021 Int. Jt. Conf. Neural Netw. IJCNN, Shenzhen, China: IEEE; 2021, p. 1–9. https://doi.org/10.1109/IJCNN52387.2021.9533606.

[60] Brewe DE. Theoretical Modeling of the Vapor Cavitation in Dynamically Loaded Journal Bearings. J Tribol 1986;108:628–37. https://doi.org/10.1115/1.3261288.